\documentclass[aps,twocolumn,preprintnumbers,amsmath,amssymb,showpacs]{revtex4}

\usepackage{amssymb}
\usepackage{amsmath}
\usepackage{graphicx}
\usepackage{bm}
\usepackage{amsfonts}

\begin{document}
\title[Self-Spin-Controlled Rotation of Spatial States of a Dirac Electron]{Self-Spin-Controlled Rotation of Spatial States of a Dirac Electron in a Cylindrical Potential via Spin-Orbit Interaction}

\author{C.C. Leary}
\email{cleary@uoregon.edu}
\author{D. Reeb}
\author{M.G. Raymer}
\affiliation{Oregon Center for Optics and Department of Physics, University of Oregon, Eugene, OR USA, 97403}

\date{\today}

\begin{abstract}
Solution of the Dirac equation predicts that when an electron with non-zero orbital angular momentum propagates in a cylindrically symmetric potential, its spin and orbital degrees of freedom interact, causing the electron's phase velocity to depend on whether its spin and orbital angular momenta vectors are oriented parallel or anti-parallel with respect to each other.  This spin-orbit splitting of the electronic dispersion curves can result in a rotation of the electron's spatial state in a manner controlled by the electron's own spin z-component value. These effects persist at non-relativistic velocities. To clarify the physical origin of this effect, we compare solutions of the Dirac equation to perturbative predictions of the Schr\"{o}dinger-Pauli equation with a spin-orbit term, using the standard Foldy-Wouthuysen Hamiltonian. This clearly shows that the origin of the effect is the familiar relativistic spin-orbit interaction. 
\end{abstract}


\maketitle

\section{\label{sec:level1}Introduction\protect}

 The physical consequences of the spin-orbit interaction (SOI) for an electron in a spherically symmetric central potential are well-known: the corrections to the bound-state eigen-energies depend on the projection of the electron's spin angular momentum (SAM) onto its orbital angular momentum (OAM), $\hat{\textbf{S}}\cdot\hat{\textbf{L}}$ \cite{Sakurai}. This energy splitting contributes to the famous fine structure of the energy states in the hydrogen atom. One can calculate it using either the exact solution of the Dirac equation \cite{Davydov,Greiner}, which includes SOI implicitly, or by perturbation theory using the Pauli-Schr\"{o}dinger equation, after explicitly adding a spin-orbit term in the Hamiltonian \cite{Shankar} proportional to $\hat{\textbf{S}}\cdot\hat{\textbf{L}}$. For an electron traveling within a cylindrically symmetric potential of infinite length, the energy states are continuous rather than discrete. However there do exist transversely bound states, and one might still expect the SOI to alter the properties of these states in some way. Surprisingly, this simple and analytically solvable problem does not seem to have been considered previously in the literature.

 In this paper we solve the problem of an electron traveling down a cylindrically symmetric step-potential that is translationally invariant in the z direction (see Fig. 1). We derive the wavefunctions and dispersion relations connecting the electron's energy and momentum. We find in the cylindrical case that the energy corrections to the transversely bound states are proportional to the product $\sigma m_{\ell}$, where $\sigma$ and $m_\ell$ are quantum numbers corresponding to the \textit{z}-components of the electron's spin vector $\hat{\mathbf{S}}$ and OAM vector $\hat{\mathbf{L}}$, respectively. This stands in contrast to the case of a central potential, where spherical symmetry dictates the dependence of the energy splittings upon quantum numbers $j$, $\ell$, and $s$ only, where $j$, $\ell$, and $s$ correspond to the electron's total angular momentum, OAM, and SAM, respectively. Similarly to the spherical case however, the cylindrical SOI arises only in the presence of an an inhomogeneous potential; this interaction is absent for an electron in free space, even when considering axially localized beam-like states. 
 
\begin{figure}
\begin{center}
\includegraphics[width=0.5\textwidth]{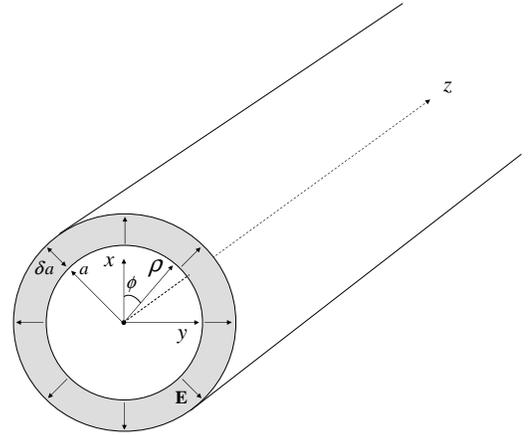}
\end{center}
\caption{\label{Fig1.eps} Two concentric cylindrical surfaces with nearly equal radii $a$ and $a+\delta a$.  The inner (outer) cylinder is positively (negatively) charged, thereby giving rise to an approximately constant electric field pointing radially outward between the cylinders, as expressed in equation \eqref{Eqn1}. The electric field is zero elsewhere.}
\end{figure}
 
 The relationship between the electron's energy and longitudinal propagation constant is given by the dispersion curves for the distinct transverse states. We calculate the splitting of the dispersion curves induced by the SOI via two methods, paralleling the two standard approaches to the spherically symmetric case discussed above. First, we employ first-order perturbation theory on the Pauli-Schr\"{o}dinger equation after explicitly adding to the Hamiltonian the appropriate spin-orbit term. In contrast to the former spherical case, we find that the added term is proportional to the product of the \textit{z}-components of the spin and OAM operators, $\hat{S}_z \hat{L}_z$. In the second approach we find nonperturbative solutions of the Dirac equation for the cylindrical geometry. The two results for the SOI splitting are found to agree in the appropriate limit, thus confirming the validity of the Hamiltonian used for the perturbative theory. 

 The splitting of the dispersion curves has the following meaning, apparently found here for the first time: for a given electron energy, the phase velocity of the electron depends on whether the quantum number $\sigma$ has equal or opposite sign as the quantum number $m_\ell$.  That is, they depend on whether $\hat{S}_{z}$ points \textit{parallel} or \textit{anti-parallel} in relation to $\hat{L}_{z}$. This coupling of $\sigma$ and $m_\ell$ has an interesting consequence: it implies that there exist stable electronic states whose transverse spatial wavefunctions \textit{rotate} as they propagate down the cylinder, with the direction of rotation depending on the sign of $\sigma$ (see Fig. 4). One can therefore in principle exploit this interaction to achieve spin-controlled manipulation of the spatial electron wavefunction.
 
 This spin-dependent rotational effect occurs in two distinct contexts (see Fig. 3): when the electron wavefunction is a superposition of degenerate energy eigenstates with the \textit{same} value of $\sigma$ but \textit{opposite} values of $m_\ell$, the rotation occurs as a function of \textit{z}.  Complementarily, when the electron is a superposition of degenerate eigenstates of the \textit{z}-component of \textit{linear momentum}, while still having the same $\sigma$ and opposite $m_\ell$, the rotation occurs as a function of \textit{time}.  The possibility of this latter type of rotation for photons was predicted in \cite{Enk07}. Both of these effects are the result of a varying relative phase between the propagating parallel and anti-parallel eigenstates, which in turn originates from the SOI-induced corrections to the dispersion mentioned above. Although these phenomena arise from relativistic dynamics, they persist even for nonrelativistic velocities.
 
 We are not aware of electron experiments to date that are sensitive to the predicted SOI effects in cylindrical geometry. Semiconductor waveguides used for studying ballistic transport of low-temperature electrons are typically rectangular in cross section, so OAM is not conserved. Electrons in linear accelerator beams do not typically have transverse coherence areas as large as the beam area, so coherent quantum effects would not be observed. In fact, the present calculation was motivated by considering the analogous problem of a single photon traveling in a cylindrical optical fiber, where analogous effects have been predicted \cite{Kapany}, \cite{Liberman}. Although in this work we consider in detail only the simple case of a step-potential, we expect the aforementioned SOI splitting effects to persist in any inhomogeneous cylindrical potential that is translationally invariant in the z direction.  However, if the requirement of translation invariance is dropped, we expect the SOI to manifest itself in a more complicated way, in analogy with predictions of SOI for photons in a cylindrical Bragg cavity \cite{Foster}. In a future paper, we will elucidate the electron-photon SOI analogy in detail.

 The remainder of this work is organized as follows: in section \ref{sec:soh}, we derive the SOI Hamiltonian using a heuristic classical model of a charged particle with a magnetic moment propagating in a cylindrical waveguide. In section \ref{sec:rotation} we quantize this Hamiltonian and employ perturbation theory, thereby deriving the aforementioned energy and propagation constant splitting, as well as the spin-controlled spatial rotation effect on the wavefunctions. After starting from the quasi-relativistic Foldy-Wouthuysen representation of the Dirac equation, we arrive in section \ref{sec:sos} at the same SOI Hamiltonian obtained in the heuristic model. We also give more explicit expressions for the first order energy and propagation constant corrections and corresponding rotation rate. We provide in section \ref{sec:dirac} the most rigorous perspective on the SOI by obtaining relativistic wavefunctions directly from the Dirac equation with a step-potential, thereby showing the equivalence of our results in the Dirac, Foldy-Wouthuysen, and heuristic pictures in the appropriate limits.  We conclude this work in section \ref{sec:conclusions} by discussing the physical origin of the SOI for electrons, and briefly comparing it to the analogous case of a photon propagating in a step-index optical fiber.

\section{\label{sec:soh}Spin-Orbit Hamiltonian\protect} 
 Consider a cylindrically symmetric potential which can be modeled by two concentric cylindrical surfaces with nearly equal radii $a$ and $a+\delta a$ (see Fig. 1). The inner cylinder is uniformly positively charged (as observed in the laboratory frame), and the outer cylinder is uniformly negatively charged, in such a way that overall the waveguide is neutral. The electric field is zero inside the inner cylinder and outside the outer cylinder, but is nonzero (and approximately constant) in the region between the cylinders, such that 

\begin{equation} \label{Eqn1} 
\mathbf{E}=\mathcal{E}_{0} \frac{a}{\rho } \Theta \left(\rho \right)\hat{\rho }\approx \mathcal{E}_{0} \Theta \left(\rho \right)\hat{\rho }
\end{equation} 

\noindent where $\Theta \left(\rho \right)\equiv \theta \left(\rho -a\right)-\theta \left(\rho -\left(a+\delta a\right)\right)$ with $\theta $ being the Heaviside step function and $\rho $ the radial distance in cylindrical coordinates, and where $\hat{\rho}$ is the radial unit vector. The approximation on the right hand side of \eqref{Eqn1} is valid in the regime where $\delta a\ll{a}$.  The magnetic field is zero everywhere in the laboratory frame.

 We are interested in the case of an electron traveling down the cylinder with magnetic moment $\vec{\mu }$ and nonzero orbital angular momentum \textit{z}-component (OAM) $L_{z} =\rho p_{\phi } $ with respect to the cylinder axis.  We also assume that the electron is moving paraxially with respect to the cylinder axis such that $\left|\mathbf{p}_{T}\right|\ll\left|\mathbf{p}_{z}\right|$, where $\mathbf{p}_{z} \equiv p_{z} \hat{\mathbf{z}}$ and $\mathbf{p}_{T} \equiv p_{\rho } \hat{\mathbf{\rho}}+p_{\phi } \hat{\mathbf{\phi}}$ are the electron's longitudinal and transverse momenta in cylindrical coordinates, respectively. We will show that when such an electron is present in the region with nonzero electric field, the electronic motion gives rise to a spin-orbit interaction between its magnetic moment \textit{z}-component $\mu _{z} $ and OAM $L_{z} $.  

 The standard theory of SOI is summarized in \cite{Rose}. The magnetic field in the (primed) rest frame of the electron is 

\begin{equation} \label{Eqn2} 
\mathbf{B}'=-\gamma \frac{\mathbf{v}}{c} \times \mathbf{E}\approx -\frac{\mathbf{v}}{c} \times \mathbf{E} 
\end{equation} 

\noindent where $v$ is the electron velocity in the laboratory frame, and the Lorentz factor $\gamma \approx 1$ for sufficiently low $v$, which we will assume throughout this section.  Also in \eqref{Eqn2}, we have employed Gaussian units, following \cite{Rose}. The presence of the electron's magnetic moment $\vec{\mu }$ in such a field gives rise to a magnetic dipole interaction energy $H'=-\vec{\mu }\cdot \mathbf{B}'$. After accounting for the relativistic Thomas precession effect \cite{RoseNote}, which effectively contributes a factor of ${\raise0.5ex\hbox{$\scriptstyle 1 $}\kern-0.1em/\kern-0.15em\lower0.25ex\hbox{$\scriptstyle 2 $}}$, this energy becomes

\begin{equation} \label{Eqn3} 
H'={\tfrac{1}{2}} \vec{\mu }\cdot \left(\frac{\mathbf{v}}{c} \times \mathbf{E} \right)=-\frac{1}{2mc} \vec{\mu }\cdot \Big( \mathbf{E}\times \left(p_{z} \hat{z}+\mathbf{p}_{T} \right) \Big) 
\end{equation} 

\noindent where $\mathbf{p}=p_{z} \hat{z}+\mathbf{p}_{T} $ is the electron momentum in the laboratory frame.  The SOI Hamiltonian therefore contains two parts in our present case with respective forms $\vec{\mu }\cdot \left(\mathbf{E}\times p_{z} \hat{\mathbf{z}}\right)$ and $\vec{\mu }\cdot \left(\mathbf{E}\times \mathbf{p}_{T} \right)$.  As the former term depends on the longitudinal momentum $p_{z}$ only, and therefore does not involve the electron's transverse OAM, we henceforth disregard it as a candidate for SOI.  Upon employing \eqref{Eqn1}, however, it is evident that the latter term involves a magnetic field vector proportional to $\mathbf{E}\times \mathbf{p}_{T} =\mathcal{E}_{0} p_{\phi } \Theta \left(\rho \right)\hat{\mathbf{z}}$, which points either \textit{parallel} or \textit{anti-parallel} with the \textit{z}-axis according to the sign of $p_{\phi } =\frac{1}{\rho } L_{z} $ (see Fig. 2). From \eqref{Eqn3}, this results in a SOI energy contribution of 

\begin{equation} \label{Eqn4} 
H_{SOI} \approx-\frac{1}{2mc} \frac{\mathcal{E}_{0} }{a } \mu _{z} L_{z} \Theta \left(\rho \right) 
\end{equation} 

\begin{figure}
\begin{center}
\includegraphics[width=0.5\textwidth]{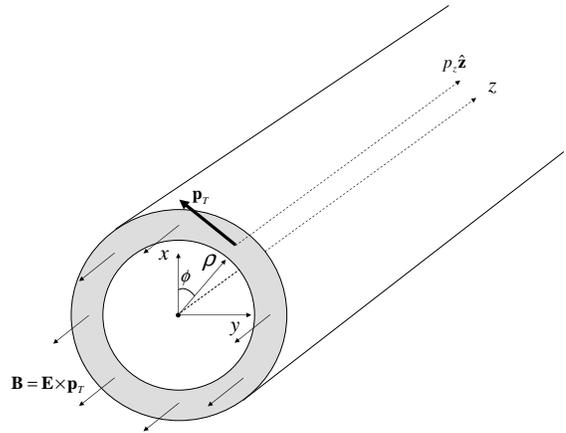}
\end{center}
\caption{\label{Fig2.eps} The magnetic field contribution due to an electron propagating paraxially between the cylinders of the waveguide with nonzero $p_{\phi }$, as experienced in the electron's rest frame. As discussed in the main text of the paper, we ignore the contribution due to $p_{z }$ (represented by the dotted arrow in the figure), so that the field shown in the figure is that due only to the transverse component of momentum $\mathbf{p}_{T}$ (represented by the bold arrow in the figure). This effective magnetic field points in the negative-\textit{z} direction for anti-clockwise $p_{\phi }$ (as shown above), and in the positive-\textit{z} direction for clockwise $p_{\phi }$.}
\end{figure}

\noindent where $ \rho \approx a $ has been used. From \eqref{Eqn4} we see that when the electron is in the region $a\le \rho \le a+\delta a$, it experiences a SOI energy shift proportional to the \textit{product }of $\mu _{z} $ and $L_{z} $.  In other words, the \textit{sign} of the spin-orbit energy shift depends upon whether $\mu _{z} $ and $L_{z} $ are pointing \textit{parallel} or \textit{anti-parallel} to each other.

\section{\label{sec:rotation}Propagation constant splitting and spin-controlled rotation\protect}

We quantize \eqref{Eqn4} by letting $\mu _{z} \to -\frac{e}{mc} \hat{S}_{z} =-\frac{e\hbar }{2mc} \hat{\sigma }_{z} $ and $L_{z} \to \frac{\hbar }{i} \frac{\partial }{\partial \phi }$ ($\hat{\sigma }_{z} $ is the Pauli matrix), so that the quantized Hamiltonian is

\begin{eqnarray}  \label{Eqn5} 
\hat{H}_{SOI}&&=\frac{e}{2m^{2}c^{2} } \frac{\mathcal{E}_{0} }{a}\hat{S}_{z} \hat{L}_{z} \Theta \left(\rho \right)\nonumber\\&&=\frac{e}{2m^2c^{2} } \frac{\mathcal{E}_{0} }{a} \left(\frac{\hbar }{2} \right)\left(\begin{array}{cc} {{\tfrac{\hbar }{i}} {\tfrac{\partial }{\partial \phi }} } & {0} \\ {0} & {-{\tfrac{\hbar }{i}} {\tfrac{\partial }{\partial \phi }} } \end{array}\right)\Theta \left(\rho \right) 
\end{eqnarray}

\noindent where $e=\left|e\right|$ is the elementary charge.

 The Hamiltonian in \eqref{Eqn5} is analogous to that which arises from an electron orbiting around a proton in a hydrogen atom---the canonical example for SOI.  In that case, the electric field can be written as $\mathbf{E}=\frac{1}{r}
 \mathcal{E}_{0}^{Coulomb} \mathbf{r}$, where $\mathcal{E}_{0}^{Coulomb}=\frac{e}{r^{2}}$ is the Coulomb field due to the proton, so that the Hamiltonian in \eqref{Eqn3} gives rise to the well-known atomic spin-orbit coupling Hamiltonian for a Coulomb potential:

\begin{equation} \label{Eqn6} 
\hat{H}_{Coulomb} =\frac{e}{2m^{2} c^{2} } \frac{\mathcal{E}_{0}^{Coulomb}}{r} \hat{\mathbf{S}}\cdot \hat{\mathbf{L}} 
\end{equation} 

\noindent Though the Hamiltonians in \eqref{Eqn5} and \eqref{Eqn6} have similar forms and in both cases the SOI arises from the same Hamiltonian \eqref{Eqn3}, the difference between the spherical and cylindrical geometries has significant physical consequences.  In particular, for the cylinder case the spin and orbital quantum angular momentum operators corresponding to the quantities $\mu _{z} $ and $L_{z} $ \textit{commute} with the Hamiltonian, while for the atomic interaction this is not the case, so that one must use the \textit{total} angular momentum operator $\hat{J}^{2}$ and the \textit{z}-component of total angular momentum $\hat{J}_{z}$ in the place of these.  Therefore, while the total angular momentum quantum numbers $j$ and $m_{j}$ are good quantum numbers for the hydrogen atom, the spin and OAM quantum numbers $\sigma$ and $m_{\ell }$ are not.  Conversely, $\sigma$, $m_{\ell}$ and $m_{j}$ are \textit{all} good quantum numbers for the cylinder case (though $j$ is not, due to the breaking of the spherical symmetry), so that states with well-defined $\sigma$ and $m_{\ell }$ are energy eigenstates.  We will make implicit use of this fact shortly.

 We treat \eqref{Eqn5} as a perturbation of the standard Schr\"{o}dinger Hamiltonian $\hat{H}_{0} =\frac{\hat{p}^{2} }{2m} -eV\left(\rho \right)$, where $V\left(\rho \right)>0$.  Our present task is therefore to find the unperturbed Schr\"{o}dinger wavefunctions.  Assuming the traveling wave form $\psi \propto e^{i\left(\beta _{0} z-\frac{E_{0} }{\hbar } t\right)} $ for the unperturbed eigenstates, in cylindrical coordinates the unperturbed equation of motion $\hat{H}_{0} \psi =E_{0} \psi $ takes the form

\begin{equation} \label{Eqn7} 
\nabla _{T}^{2} \psi +\kappa _{0}^{2} \psi =0 
\end{equation}

\noindent where \textbf{$\nabla _{T}^{2} $ }is the transverse Laplacian $\nabla ^{2} -{\tfrac{\partial ^{2} }{\partial z^{2} }} $, and the transverse wavenumber is

\begin{equation} \label{Eqn8} 
\kappa _{0}^{2} \equiv \frac{2m}{\hbar ^{2} } \left(E_{0} +eV\left(\rho \right)\right)-\beta _{0}^{2}  
\end{equation} 

\noindent For a constant electric potential $V\left(\rho \right)=V_{0} $ inside the cylinder, this is Bessel's equation, with solutions

\begin{equation} \label{Eqn9} 
{\left| \psi _{0}  \right\rangle} =NJ_{\left|m_{\ell } \right|} \left(\kappa _{0} \rho \right)e^{im_{\ell } \phi } \left(\begin{array}{c} {\delta _{\sigma +} } \\ {\delta _{\sigma -} } \end{array}\right)e^{i\left(\beta _{0} z-\frac{E_{0} }{\hbar } t\right)}  
\end{equation} 

\noindent where we have constrained the wavefunctions to be finite at the origin. In \eqref{Eqn9}, $N$ is a normalization constant, the radial function $J_{\left|m_{\ell}\right|} \left(\kappa _{0} \rho \right)$ is a Bessel function of the first kind of order $\left|m_{\ell}\right|=0,1,2,...$, and $\left(\begin{array}{c} {\delta _{\sigma +} } \\ {\delta _{\sigma -} } \end{array}\right)$ is a two component spinor composed of Kronecker delta functions such that $\delta _{\sigma +} =1$ if  $\sigma =+1$ and $\delta _{\sigma +} =0$ if  $\sigma =-1$, etc.  In expressing these wavefunctions, we have chosen the following complete set of commuting operators, $\{ \hat{H},\hat{p}_{z} ,\hat{L}_{z} ,\hat{S}_{z} \} $, which have the following respective eigenvalues, $\{ E_{0} ,\hbar \beta _{0} ,\hbar m_{\ell } ,{\tfrac{\hbar }{2}} \sigma \} $. We will henceforth designate the states in \eqref{Eqn9} by ${\left| \psi _{0}  \right\rangle} \equiv {\left| m_{\ell } ,\sigma  \right\rangle} $. 

 From \eqref{Eqn5} and \eqref{Eqn9} we conclude that the first-order correction to the energy of an unperturbed state, $\delta E={\left\langle m_{\ell } ,\sigma  \right|} \hat{H}_{SOI} {\left| m_{\ell } ,\sigma  \right\rangle} $, is proportional to product $\sigma m_{\ell } $ provided that the wavefunction is nonzero in the region $a\le \rho \le a+\delta a$.  This is indeed always the case for the transverse bound electronic states in \eqref{Eqn9} (we will show this in section \ref{sec:sos} when we apply the appropriate boundary conditions).  Explicitly, the first-order energy shift in this heuristic model is

\begin{align} \label{Eqn10} 
\delta E_{\sigma m_{\ell } } &= \sigma m_{\ell }  \frac{\mathcal{E}_{0}e\hbar^{2}}{4m^{2}c^{2} } \left\{2\pi N^{2} \frac{1}{a} \int _{a}^{a+\delta a}\rho d\rho J_{\left|m_{\ell } \right|}^{2} \left(\kappa _{0} \rho \right)  \right\} \nonumber \\
&\approx \sigma m_{\ell } \frac{\mathcal{E}_{0}e \pi \hbar ^{2} \delta a}{2m^{2} c^{2} } N^{2} J_{\left|m_{\ell } \right|}^{2} \left(\kappa _{0} a\right)
\end{align} 

\noindent Therefore, if the electron's SAM points \textit{parallel} to its OAM $m_{\ell } $, then the energy will shift \textit{upward}, while for the \textit{anti-parallel }case the shift will be \textit{downward}.  

 As introduced previously, two physical consequences of \eqref{Eqn10} are the splitting of the phase velocity (and therefore also the propagation constant $\beta_{0}$) of electron cylinder wavefunctions with different values of $\sigma m_{\ell } $, and the related spin-controlled spatial rotation of these wavefunctions. In order to better understand these effects, we note that due to the electron's wavelike properties, we can think of \eqref{Eqn8} as a dispersion relation defining $\beta \left(E\right)=\beta \left(\hbar \omega \right)$:

\begin{equation} \label{Eqn11} 
\beta \left(\hbar \omega \right)=\sqrt{\frac{2m}{\hbar ^{2} } \left(\hbar \omega +eV_{0} \right)-\kappa ^{2} }  
\end{equation} 

\noindent Later, in section \ref{sec:dirac}, we show that the Dirac boundary conditions imply in general that the value for an electron's transverse wavenumber $\kappa $ \textit{differs} slightly according to whether $\sigma m_{\ell }$ is positive or negative (that is, whether $S_z$ and $L_z$ are parallel or anti-parallel). We thus employ positive and negative sign superscripts to denote these two cases, so that $\kappa \to \kappa ^{+} $ or $\kappa \to \kappa ^{-} $ depending on whether the the SAM and OAM are parallel or anti-parallel, etc.  Therefore, we conclude from \eqref{Eqn11} that a parallel and anti-parallel state with the \textit{same} value for $\beta $ will have slightly \textit{differing} frequency (energy) values $\omega ^{+} $ and $\omega ^{-} $, respectively.  This is the energy splitting which we have calculated in \eqref{Eqn10}. However, we can also use \eqref{Eqn11} to argue the converse---that parallel and anti-parallel states with the \textit{same} frequency (energy) value $\omega $ will have slightly \textit{differing} values for their propagation constants $\beta ^{+} $ and $\beta ^{-} $, respectively. 

 For a visualization of this point, refer to Fig. 3, which gives a plot of the dispersion relations $\beta\left(E\right)$ for the states with $\sigma m_{\ell }=+1$ and $\sigma m_{\ell }=-1$, thereby explicitly showing the splitting of the curves (the dotted curve is a plot of the unperturbed dispersion relation). These parallel and anti-parallel states have different energies $E_{\parallel}$ ($E^{+}$) and $E_{\not\parallel}$ ($E^{-}$) for a fixed value for the propagation constant $\beta_{0}$, as shown by the solid vertical and horizontal lines in the figure. Conversely, the two states have different $\beta$ values $\beta_{\parallel}$ ($\beta^{+}$) and $\beta_{\not\parallel}$ ($\beta^{-}$) for a fixed value of the \textit{energy} $E_{0}$, as shown by the \textit{dotted} vertical and horizontal lines. The horizontal and vertical arrows respectively show the directions (signs) of the energy and propagation constant shifts $\delta E$ and $\delta \beta$ for a parallel state (for an anti-parallel state, the signs of both $\delta E$ and $\delta \beta$ are switched). The inlaid picture shows the resultant transverse spatial probability density distribution when the parallel (see equation \eqref{Eqn12a}) and anti-parallel (see equation \eqref{Eqn12b}) states with $|m_{\ell}|=1$ are superposed, as given by equation \eqref{Eqn14}. 

\begin{figure}
\begin{center}
\includegraphics[width=0.5\textwidth]{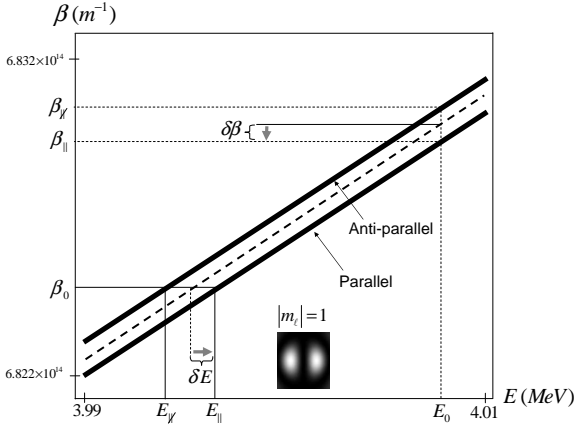}
\end{center}
\caption{\label{Fig3.eps} The splitting of the parallel and anti-parallel states involving $m=1$. This plot is a blown-up version of the inlaid box in Fig. 4, which plots the dispersion curves for all allowed states ${\left| m_{\ell } ,\sigma  \right\rangle}$ with $R=6$ and $e \Delta V=0.02mc^{2}$ (for definitions of $R$ and $\Delta V$, see equations \eqref{Eqn23} and \eqref{Eqn16}, respectively). The dashed curve is a plot of the unperturbed dispersion relation. The inlaid picture is a plot of the resulting transverse spatial probability density distribution when the parallel and anti-parallel states with $|m_{\ell}|=1$ are superposed. For a fixed $\beta$, the azimuthal lobes of this distribution rotate as a function of time, while for fixed energy they rotate as a function of distance down the cylinder as shown in equation \eqref{Eqn14}. In both cases the direction of rotation is dependent upon the spin of the superposition mode. For further discussion and the interpretation of the intersecting vertical and horizontal lines and arrows, see the main text of the paper.}
\end{figure}

In order to calculate the propagation constant shift $\delta \beta $ to first order in terms of the energy shift $\delta E=\hbar \delta \omega $ which we have already found, we expand the propagation constant $\beta \left(\hbar \omega \right)$ to first order in $\hbar \omega $ about the unperturbed energy value $\hbar \omega _{0} $, thus approximating $\beta \left(\hbar \omega \right)\approx \beta \left(\hbar \omega _{0} \right)\pm |\delta \beta| $.  In this way, the absolute value $|\delta \beta|$ can be written as

\begin{equation} \label{Eqn11b} 
|\delta \beta _{\sigma m_{\ell } }| =\left|\frac{\partial \beta }{\hbar \partial \omega } \left|_{\omega =\omega _{0} } \right. \hbar \left(\omega -\omega _{0} \right)\right|=\frac{1}{\hbar v_{z} \left(\omega _{0} \right)} |\delta E_{\sigma m_{\ell } }|  
\end{equation} 

\noindent where $\delta E_{\sigma m_{\ell } } $ is the first-order energy shift, and $v_{z} \left(\omega _{0} \right)$ is the (positive) \textit{z-}component of the group velocity of the matter wave, which is interpreted as the velocity of the electron as it travels down the cylinder. In order to determine the relative sign of $\delta \beta _{\sigma m_{\ell } }$ and $\delta E_{\sigma m_{\ell } }$, refer to Fig. 3. From the figure it is evident that for the dispersion curve of the parallel state, the energy shift $\delta E_{\sigma m_{\ell } } $ is positive while the propagation constant shift $\delta \beta _{\sigma m_{\ell } }$ is negative, as shown by the horizontal and vertical arrows. Conversely, for the dispersion curve of the anti-parallel state, $\delta E_{\sigma m_{\ell } } $ is negative while $\delta \beta _{\sigma m_{\ell } }$ is positive. We therefore conclude that 

\begin{equation} \label{Eqn12} 
\delta \beta _{\sigma m_{\ell } } =-\frac{1}{\hbar v_{z} \left(\omega _{0} \right)} \delta E_{\sigma m_{\ell } }  
\end{equation}

 This splitting in the propagation constants between electrons with parallel and anti-parallel spin and orbital angular momenta has a remarkable consequence: If one superposes a parallel and anti-parallel state with the \textit{same} value for $\sigma $ and the same \textit{absolute} value for $m_{\ell}$, then the orbital angular momentum of the \textit{parallel} state will be $\sigma \left|m_{\ell } \right|$, while the orbital angular momentum of the \textit{anti-parallel} state will be $-\sigma \left|m_{\ell } \right|$.  In the quasi-paraxial regime the resulting superposition wavefunction will then possess an azimuthal pattern that \textit{rotates }as the particle propagates in the step-potential, with the sense of the rotation depending on the spin $\sigma $. This spin-controlled rotation effect is a direct result of the varying relative phase between the parallel and anti-parallel states as they propagate down the cylindrical potential, which is in turn caused by the difference in the propagation constants of these states. 

More concretely, from \eqref{Eqn9}, a \textit{parallel} state has the approximate form \footnote{Strictly speaking, the wave numbers $\kappa^{+}$ and $\kappa^{-}$ of the parallel and anti-parallel states are not equal to each other or to $\kappa_{0}$ as given in \eqref{Eqn12a} and \eqref{Eqn12b}. However, the approximation $\kappa^{+}\approx\kappa^{-}\approx\kappa^{0}$ is indeed justified as we show near the end of section \ref{sec:dirac} via solution of the Dirac equation.}

\begin{equation} \label{Eqn12a} 
{\left| \psi _{\parallel}  \right\rangle} =NJ_{\left|m_{\ell } \right|} \left(\kappa _{0} \rho \right)\left(\begin{array}{c} {\delta _{\sigma +} } \\ {\delta _{\sigma -} } \end{array}\right)e^{i\left(\sigma \left|m_{\ell } \right|\phi +\beta ^{+} z\right)} e^{-i\frac{E_{0} }{\hbar } t}
\end{equation}

\noindent inside the cylinder, while an \textit{anti-parallel} state is 

\begin{equation} \label{Eqn12b} 
{\left| \psi _{\not\parallel}  \right\rangle} =NJ_{\left|m_{\ell } \right|} \left(\kappa _{0} \rho \right)\left(\begin{array}{c} {\delta _{\sigma +} } \\ {\delta _{\sigma -} } \end{array}\right)e^{-i\left(\sigma \left|m_{\ell } \right|\phi -\beta ^{-} z\right)} e^{-i\frac{E_{0} }{\hbar } t} 
\end{equation}

\begin{figure}
\begin{center}
\includegraphics[width=0.5\textwidth]{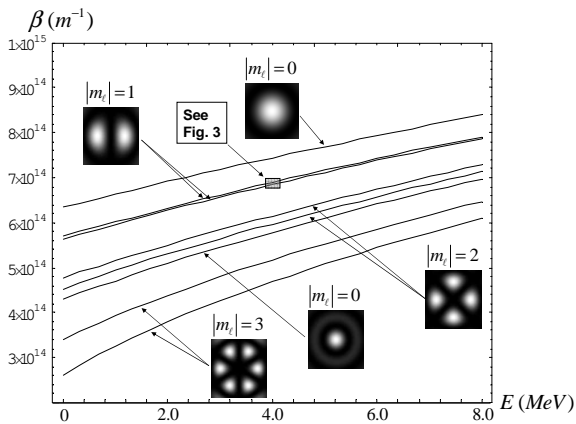}
\end{center}
\caption{\label{Fig4.eps} Dispersion curves for each of the allowed transversely bound electronic states ${\left| m_{\ell } ,\sigma  \right\rangle}$ for $R=6$ and $eV=0.02mc^{2}$. Note the splittings of the parallel and anti-parallel curves for states with $|m_{\ell}|\neq0$, which have been exaggerated by a factor of 50 with respect to their actual values for purposes of visualization. The inlaid pictures are electronic transverse spatial probability density distributions associated with various values of $|m_{\ell}|$, as discussed in the main text below. The varying azimuthal lobe structure of each of the plotted superposition states undergoes clockwise or counter-clockwise spin-controlled rotation as shown in equation \eqref{Eqn14}. A blown up plot of the small inlaid box which intersects the curves with $|m_{\ell}|=1$ is presented as Fig. 3.}
\end{figure}

\noindent The equal superposition of these two states, which we denote as ${\left| \psi _{\sigma }  \right\rangle} $, is therefore equal to

\begin{align} \label{Eqn13} 
{\left| \psi _{\sigma }  \right\rangle} =N&J_{\left|m_{\ell } \right|} \left(\kappa _{0} \rho \right)\left(\begin{array}{c} {\delta _{\sigma +} } \\ {\delta _{\sigma -} } \end{array}\right)\nonumber \\ &\times \left(e^{i\left(\sigma \left|m_{\ell } \right|\phi +\beta ^{+} z\right)}
+e^{-i\left(\sigma \left|m_{\ell } \right|\phi -\beta ^{-} z\right)} \right)e^{-i\frac{E_{0} }{\hbar } t}  
\end{align} 

\noindent Note however that $\left(e^{i\left(\sigma \left|m_{\ell } \right|\phi +\beta ^{+} z\right)} +e^{-i\left(\sigma \left|m_{\ell } \right|\phi -\beta ^{-} z\right)} \right)$ can be written as $\cos \left(\left|m_{\ell } \right|\phi +\sigma \Delta \beta z\right)e^{i\bar{\beta }z} $, where $\Delta \beta \equiv {\tfrac{1}{2}} \left(\beta ^{+} -\beta ^{-} \right)$ and $\bar{\beta }\equiv {\tfrac{1}{2}} \left(\beta ^{+} +\beta ^{-} \right)$ thereby leading us to our final result,

\begin{align} \label{Eqn14} 
{\left| \psi _{\sigma }  \right\rangle} &= \nonumber \\ & NJ_{\left|m_{\ell } \right|} \left(\kappa _{0} \rho \right)\cos \left(\left|m_{\ell } \right|\phi +\sigma \Delta \beta z\right)\left(\begin{array}{c} {\delta _{\sigma +} } \\ {\delta _{\sigma -} } \end{array}\right)e^{i\left(\bar{\beta }z-\frac{E_{0} }{\hbar } t\right)}  
\end{align} 

 Equation \eqref{Eqn14} is a major result of this paper; for a cylindrical step-potential it predicts the existence of Schr\"{o}dinger wavefunctions with an azimuthal lobe structure that rotate clockwise or counterclockwise about the cylinder axis as the particle propagates, with the sense of the rotation depending on the spin $\sigma$. Furthermore, the rotation rate $\Delta \beta $ of the wavefunction has already been given implicitly via \eqref{Eqn10} and \eqref{Eqn12}, and will be calculated explicitly using two different approaches in sections \ref{sec:sos} and \ref{sec:dirac}.

 Dispersion curves associated with the parallel and anti-parallel states in equations \eqref{Eqn12a} and \eqref{Eqn12b} are plotted in Fig. 4 for several values of $m_{\ell}$. The inlaid pictures associated with each value of $|m_{\ell}|$ are plots of electronic transverse spatial probability density distributions. The distributions labeled by $|m_{\ell}|=0$ are obtained via equation \eqref{Eqn9}, while those labeled by $|m_{\ell}|=1,2,$ and $3$ are the rotating superposition states as given by \eqref{Eqn14}, which result from superposing the parallel and anti-parallel states with equal $|m_{\ell}|$ from equations \eqref{Eqn12a} and \eqref{Eqn12b}. For a superposition state of a given energy, the spatial rotation rate $\Delta\beta$ is just half the splitting between its associated parallel and anti-parallel dispersion curves. The method used for calculating the curves is derived presently in section \ref{sec:sos}: for a given value of $|m_{\ell}|$ we solve \eqref{Eqn22} and \eqref{Eqn23} for $\kappa_0$ and then substitute the result into equation \eqref{Eqn11}. For the values chosen for the figure, there are two allowed solutions for $|m_{\ell}|=0$, giving rise to two distinct $|m_{\ell}|=0$ dispersion curves and therefore also two distinct probability densities. For $|m_{\ell}|=1,2,$ and $3$, the dispersion curve splitting gives rise to stable superposition states as shown.

\section{\label{sec:sos}Spin-Orbit Shift: Explicit Calculation\protect}

 In order to obtain explicit results for the SOI energy and propagation constant shifts, it is instructive to approach the problem from the more rigorous viewpoint of the Foldy-Wouthuysen representation \cite{Greiner}, wherein the Dirac Hamiltonian has the general property that the positive energy solutions are decoupled from the negative energy solutions so that we can describe the electron via a two-component spinor. In the presence of an arbitrary electrostatic field in the laboratory frame (the magnetic field is zero), to order $\left({\tfrac{v}{c}} \right)^{4} $, the Dirac Hamiltonian in the Foldy-Wouthuysen representation takes the form \cite{Greiner}

\begin{align} \label{Eqn15} 
\hat{H}_{\Phi } =\frac{1}{2m} \hat{p}^{2} -eV\left(r\right) & +\left\{-\frac{1}{8m^{3} c^{6} } \hat{p}^{4} +\frac{ie\hbar }{4m^{2} c^{2} } \hat{\mathbf{S}}\cdot \left(\nabla \times \mathbf{E}\right) \right. \nonumber \\ 
& \left. +\frac{e}{2m^{2} c^{2} } \hat{\mathbf{S}}\cdot \left(\mathbf{E}\times \hat{\mathbf{p}}\right)+\frac{e\hbar ^{2} }{8m^{2} c^{2} } \nabla \cdot \mathbf{E} \right\}
\end{align} 

\noindent where $\hat{\mathbf{S}}$ is the spin vector operator of $2\times2$ Pauli matrices, the rest mass term has been dropped, and Gaussian units have again been employed.  Our first goal is to argue that the contribution of the terms in curly brackets to the SOI has the form of the heuristically derived equation \eqref{Eqn5}.  Note that the first term in curly brackets arises from the relativistic mass increase, and is independent of the form of the electric field $\mathbf{E}$.  In the canonical case of a Coulomb field, the next two terms (which are only Hermitian when taken together) give rise to the atomic spin-orbit interaction, while the last term becomes the well-known Darwin term.  In light of this, we expect only the two middle terms to contribute to the SOI in the cylindrical case, and we henceforth drop the first and fourth terms.  In section \ref{sec:dirac} we show that this is indeed justified by comparing the results of this section to those obtained directly from the Dirac equation.

 For electrostatic fields with zero curl the second term in the curly brackets also vanishes, so that after dropping the aforementioned terms there remains only the term $\frac{e}{2m^{2} c^{2} } \hat{\mathbf{S}}\cdot \left(\mathbf{E}\times \hat{\mathbf{p}}\right)$. Furthermore, since $\mathbf{E}=-\nabla V$, for a translationally invariant cylindrically symmetric potential $V\left(\rho\right)$ this term becomes
 
\begin{equation} \label{Eqn15b}
-\frac{e}{2m^{2} c^{2}}\left(\frac{1}{\rho}\frac{\partial V\left(\rho\right)}{\partial\rho}\right) \hat{\mathbf{S}}\cdot \left( \vec{\rho}\times \hat{\mathbf{p}} \right)
\end{equation}

\noindent (for a spherically symmetric atomic potential, this spin-orbit term has the same form, but with the replacement $\rho\rightarrow r$) \cite{Rose}. From \eqref{Eqn15b} it is clear from the derivative that the spin-orbit interaction depends on the inhomogeneity of the potential and therefore does not occur in free space, thereby confirming the corresponding statements made in the Introduction.

 We now introduce the cylindrical step-potential

\begin{equation} \label{Eqn16} 
V\left(\rho \right)=V_{0} -\Delta V\theta \left(\rho -a\right) 
\end{equation} 

\noindent where $V_{0} >0$ and $\Delta V>0$. Substituting \eqref{Eqn16} into \eqref{Eqn15b}, and noting that $\frac{\partial \theta \left(\rho -a\right)}{\partial \rho}=\delta\left(\rho -a\right)$ (where $\delta\left(\rho\right)$ denotes the Dirac delta function), we find that the Hamiltonian in \eqref{Eqn15} takes the form (after dropping the aforementioned terms in curly brackets)

\begin{equation} \label{Eqn18} 
\hat{H}_{\Phi } =\hat{H}_{0} +\hat{H}_{SOI}  
\end{equation}

\noindent where $\hat{H}_{0} =\frac{\hat{p}^{2}}{2m} -eV\left(\rho \right)$ is the standard Schr\"{o}dinger Hamiltonian and

\begin{equation} \label{Eqn19} 
\hat{H}_{SOI} =\frac{e}{2m^{2} c^{2} } \frac{\Delta V}{\rho } \delta \left(\rho -a\right)\hat{\mathbf{S}}\cdot \left( \vec{\rho }\times \hat{\mathbf{p}} \right) 
\end{equation} 

\noindent is the perturbative SOI Hamiltonian.  Furthermore, note that $\hat{\mathbf{S}}\cdot \left(\vec{\rho }\times \hat{\mathbf{p}}\right)$ in $\hat{H}_{SOI} $ can be expressed as $\hat{S}_{z} \hat{L}_{z} +\left(y\hat{S}_{x} -x\hat{S}_{y} \right)\hat{p}_{z} $.  Since the unperturbed eigenstates ${\left| m_{\ell } ,\sigma  \right\rangle} $ of $\hat{H}_{0} $ have already been given via \eqref{Eqn8} and \eqref{Eqn9}, we focus on the expectation of $\hat{H}_{SOI} $ as expressed in the unperturbed state basis, which is thereby proportional to the following two terms:

\begin{align} \label{Eqn19b}
\left\langle H_{SOI} \right\rangle \propto &{\left\langle m'_{\ell } ,\sigma ' \right|} \hat{S}_{z} \hat{L}_{z} {\left| m_{\ell } ,\sigma  \right\rangle} \nonumber \\ &+{\left\langle m'_{\ell } ,\sigma ' \right|} \left(y\hat{S}_{x} -x\hat{S}_{y} \right)\hat{p}_{z} {\left| m_{\ell } ,\sigma  \right\rangle}
\end{align}

 However, since ${\left\langle m'_{\ell } ,\sigma ' \right|} \left(y\hat{S}_{x} -x\hat{S}_{y} \right)\hat{p}_{z} {\left| m_{\ell } ,\sigma  \right\rangle} $ always vanishes, we conclude that for the purposes of first-order perturbation theory we can write 

\begin{equation} \label{Eqn20} 
H_{SOI} =\frac{e}{2m^{2} c^{2} } \frac{\Delta V}{\rho } \delta \left(\rho -a\right)\hat{S}_{z} \hat{L}_{z}  
\end{equation} 

\noindent Note that this is equivalent to our dropping of the term proportional to $\vec{\mu }\cdot \left(\mathbf{E}\times p_{z} \hat{z}\right)$ in \eqref{Eqn3}.  The operator $\hat{H}_{SOI} $ in \eqref{Eqn20} is diagonal in the unperturbed basis, so we can readily calculate the energy shifts of the unperturbed eigenstates,

\begin{equation} \label{Eqn21} 
\delta E_{\sigma m_{\ell } } =\sigma m_{\ell } \frac{\pi \hbar ^{2} e\Delta V}{2m^{2} c^{2} } N^{2} J_{\left|m_{\ell } \right|}^{2} \left(\kappa _{0} a\right) 
\end{equation} 

\noindent which agrees with the heuristically derived equation \eqref{Eqn10}, since $\Delta V\approx E_{0} \delta a$.  

 Though we have managed to obtain the general form of the energy shifts without considering the boundary conditions, we must do so now in order to obtain explicit numerical results. We have already required that the wavefunctions be finite at the origin, resulting in \eqref{Eqn9}, which is valid inside the cylinder. In addition to this, we furthermore constrain ${\left| \psi _{0}  \right\rangle} $ to be zero at infinity, with both ${\left| \psi _{0}  \right\rangle} $ and its derivative continuous at the boundary (where $\rho=a$). For the region outside the cylinder, the former condition results in the modification of \eqref{Eqn9} via the replacement $J_{\left|m_{\ell } \right|} \left(\kappa _{0} \rho \right) \rightarrow K_{\left|m_{\ell } \right|} \left(\tilde{\kappa}_{0} \rho \right)$, where $K_{\left|m_{\ell } \right|} \left(\tilde{\kappa}_{0} \rho \right)$ is a modified Bessel function of the second kind of order $\left|m_{\ell } \right|$ ($\kappa _{0}$ and $\tilde{\kappa }_{0}$ denote the values of the transverse wavenumber inside and outside the boundary, respectively, as defined through \eqref{Eqn8}). After employing the well-known cylinder function recursion relations \cite{TableofIntegrals}, the latter two conditions thereby lead to the characteristic equation 
 
\begin{equation} \label{Eqn22} 
\kappa _{0} a\frac{J_{\left|m_{\ell } \right|+1} \left(\kappa _{0} a\right)}{J_{\left|m_{\ell } \right|} \left(\kappa _{0} a\right)} =\tilde{\kappa }_{0} a\frac{K_{\left|m_{\ell } \right|+1} \left(\tilde{\kappa }_{0} a\right)}{K_{\left|m_{\ell } \right|} \left(\tilde{\kappa }_{0} a\right)}  
\end{equation} 

\noindent Equation \eqref{Eqn22} is an equation in the two unknowns $\kappa _{0}$ and $\tilde{\kappa }_{0}$; in order to find a second equation in these variables, we use \eqref{Eqn16} to evaluate \eqref{Eqn8} inside and outside the cylinder and subtract the results to obtain

\begin{equation} \label{Eqn23} 
\left(\tilde{\kappa }_{0}^{2} -\kappa _{0}^{2} \right)a^{2} =2\left(\frac{2\pi a}{\lambda } \right)^{2} \left(\frac{e\Delta V}{mc^{2} } \right)\equiv R^{2}  
\end{equation} 
 
\noindent where $\lambda \equiv h/mc $ is the electron's Compton wavelength ($h$ is Planck's constant).  Equations \eqref{Eqn22} and \eqref{Eqn23} can be simultaneously solved for $\kappa _{0} $, and the result substituted into \eqref{Eqn9}, which allows us to conclude that the wavefunction is indeed nonzero at the boundary as required in section \ref{sec:soh}.  Finally, from \eqref{Eqn9}, the normalization factor in \eqref{Eqn21} is found to be

\begin{align} \label{Eqn24} 
N^{2} =\frac{1}{\pi a^{2} } \frac{1}{J_{\left|m_{\ell } \right|}^{2} \left(\kappa _{0} a\right)} & \left\{\frac{K_{\left|m_{\ell } \right|-1} \left(\tilde{\kappa }_{0} a\right)K_{\left|m_{\ell } \right|+1} \left(\tilde{\kappa }_{0} a\right)}{K_{\left|m_{\ell } \right|}^{2} \left(\tilde{\kappa }_{0} a\right)} \right. \nonumber \\ & \left.-\frac{J_{\left|m_{\ell } \right|-1} \left(\kappa _{0} a\right)J_{\left|m_{\ell } \right|+1} \left(\kappa _{0} a\right)}{J_{\left|m_{\ell } \right|}^{2} \left(\kappa _{0} a\right)} \right\}^{-1}  
\end{align} 

\noindent Therefore, \eqref{Eqn21} and \eqref{Eqn12} give the propagation constant corrections as

\begin{align} \label{Eqn25} 
\delta \beta _{\sigma m_{\ell } } =-\sigma m_{\ell } \frac{1}{v_{z} } \frac{\hbar e\Delta V}{2m^{2} c^{2} a^{2} } & \left\{\frac{K_{\left|m_{\ell } \right|-1} \left(\tilde{\kappa }_{0} a\right)K_{\left|m_{\ell } \right|+1} \left(\tilde{\kappa }_{0} a\right)}{K_{\left|m_{\ell } \right|}^{2} \left(\tilde{\kappa }_{0} a\right)}\right. \nonumber \\ & \left. -\frac{J_{\left|m_{\ell } \right|-1} \left(\kappa _{0} a\right)J_{\left|m_{\ell } \right|+1} \left(\kappa _{0} a\right)}{J_{\left|m_{\ell } \right|}^{2} \left(\kappa _{0} a\right)} \right\}^{-1}  
\end{align} 

\noindent so that $ \Delta \beta={\tfrac{1}{2}} \left(\delta \beta _{\parallel} -\delta \beta _{\not\parallel} \right)={\tfrac{1}{2}} \left(\delta \beta _{+\left|m_{\ell } \right|} -\delta \beta _{-\left|m_{\ell } \right|} \right) $ can be written as

\begin{align} \label{Eqn26} 
\Delta \beta =-\left|m_{\ell } \right|\frac{1}{v_{z} } \frac{\hbar e\Delta V}{2m^{2} c^{2} a^{2} } & \left\{\frac{K_{\left|m_{\ell } \right|-1} \left(\tilde{\kappa }_{0} a\right)K_{\left|m_{\ell } \right|+1} \left(\tilde{\kappa }_{0} a\right)}{K_{\left|m_{\ell } \right|}^{2} \left(\tilde{\kappa }_{0} a\right)}\right. \nonumber \\ & \left. -\frac{J_{\left|m_{\ell } \right|-1} \left(\kappa _{0} a\right)J_{\left|m_{\ell } \right|+1} \left(\kappa _{0} a\right)}{J_{\left|m_{\ell } \right|}^{2} \left(\kappa _{0} a\right)} \right\}^{-1}  
\end{align} 

\noindent This is the explicit form for the rotation rate of the electron spatial wavefunction as defined in equation \eqref{Eqn14}.

\section{\label{sec:dirac}Dirac equation solutions\protect}

A few gaps persist so far in the development of this work.  Specifically, in section \ref{sec:rotation} we relied on the result that the electron's transverse wavenumber $\kappa $ \textit{differs} slightly according to whether $\sigma m_{\ell } $ is positive or negative, while in section \ref{sec:sos} we assumed that neither the relativistic mass increase nor the Darwin term contributes to the SOI.  Also, we have implicitly assumed throughout the validity of the paraxial approximation, which is expressed as $\left|\mathbf{p}_{T} \right|\ll\left|\mathbf{p}_{z} \right|$, or equivalently as $\kappa\ll\beta $. In this section we will demonstrate the validity of each of these assumptions by deriving the relativistic analogue of equations \eqref{Eqn14} and \eqref{Eqn26}, obtaining the bispinorial wavefunctions directly from the Dirac equation.  Our derivation involves several steps.  First, we construct the wavefunctions of interest, and boost them to a convenient frame.  Next we apply appropriate boundary conditions and derive a characteristic equation.  Finally, we approximate this equation to the appropriate order, thereby showing its equivalence to result \eqref{Eqn26} in section \ref{sec:sos}.

 The Dirac equation in bispinor form for an electron in a constant electric potential $V\left(\rho \right)=V_{0} >0$ is

\begin{equation} \label{Eqn27} 
\left(\begin{array}{cc} {mc^{2} } & {c\vec{\sigma} \cdot \mathbf{p}} \\ {c\vec{\sigma} \cdot \mathbf{p}} & {-mc^{2} } \end{array}\right)\left(\begin{array}{c} {\chi ^{+} } \\ {\chi ^{-} } \end{array}\right)=\left(\pm E+eV_{0} \right)\left(\begin{array}{c} {\chi ^{+} } \\ {\chi ^{-} } \end{array}\right) 
\end{equation} 

\noindent where $E>0$ is the absolute energy of the particle, and the upper and lower signs correspond to positive and negative energy solutions, respectively.  Free space solutions to the Dirac equation in cylindrical coordinates have been found \cite{Balantekin}.  Since the potential $V\left(\rho \right)$ is piecewise-constant for our case of interest, the solutions to \eqref{Eqn27} will have the same form (before boundary matching) as the ones in \cite{Balantekin}.  Following \cite{Balantekin}, we choose a complete set of commuting operators as $\{ \hat{H},\hat{p}_{T} ,\hat{J}_{z} ,\hat{h}_{T} \} $, with corresponding eigenvalues $\{ \pm E,\hbar \kappa ,\hbar m_{j} ,\hbar \sigma _{T} \} $, where $\hat{J}_{z} =\hat{L}_{z} +\hat{S}_{z} $ is the total angular momentum operator and $\hat{h}_{T} =\gamma _{5} \gamma _{3} \frac{\Sigma \cdot p_{T} }{\left|p_{T} \right|} $ is the transverse helicity operator with $\Sigma \equiv\left(\begin{array}{cccc} {1} & {} & {} & {} \\ {} & {-1} & {} & {} \\ {} & {} & {1} & {} \\ {} & {} & {} & {-1} \end{array}\right)$ such that its eigenvalue $\sigma _{T} =\pm 1$, while $m_{j} $ is half an odd integer.

For simplicity (and in order to avoid Klein's paradox as discussed below), we will focus on the positive energy solutions to \eqref{Eqn27}, which are of the form

\begin{equation} \label{Eqn28} 
{\left| E,\kappa ,m_{j} ,\pm 1 \right\rangle} \equiv \left(\begin{array}{c} {\chi _{m_{j} }^{\pm } } \\ {\frac{\hbar c\left(\beta \mp i\kappa \right)}{mc^{2} +E+eV_{0} } \chi _{m_{j} }^{\mp } } \end{array}\right)e^{i\left(\beta z-\frac{E}{\hbar } t\right)}  
\end{equation} 

\noindent where 

\begin{equation} \label{Eqn29} 
\chi _{m_{j} }^{\pm } \equiv \left(\begin{array}{c} {Z_{m_{j} -{\tfrac{1}{2}} } \left(\kappa \rho \right)e^{i\left(m_{j} -{\tfrac{1}{2}} \right)\phi } } \\ {\pm Z_{m_{j} +{\tfrac{1}{2}} } \left(\kappa \rho \right)e^{i\left(m_{j} +{\tfrac{1}{2}} \right)\phi } } \end{array}\right) 
\end{equation} 

\noindent and $Z_{n} \left(\kappa _{0} \rho \right)$ denotes an arbitrary cylinder function of order $n$. Equation \eqref{Eqn27} contains the relativistic analogue of \eqref{Eqn8},

\begin{equation} \label{Eqn30} 
\left(cp\right)^{2} =\hbar ^{2} c^{2} \left(\beta ^{2} +\kappa ^{2} \right)=\left(E+eV\left(\rho \right)\right)^{2} -m^{2} c^{4}  
\end{equation} 

\noindent which for a sufficiently small step-potential \footnote{We use a small potential satisfying $eV_{0}<<mc^{2}$ in order to avoid the violation of particle number conservation, called Klein's Paradox. See also the discussion above equation \eqref{Eqn34}.}, can be used to derive a relativistic analogue to \eqref{Eqn23},

\begin{equation} \label{Eqn31} 
\left(\tilde{\kappa }^{2} -\kappa ^{2} \right)a^{2} =2\gamma \left(\frac{2\pi a}{\lambda } \right)^{2} \left(\frac{e\Delta V}{mc^{2} } \right)\equiv R_{\gamma }^{2}  
\end{equation} 

\noindent where $E=\gamma mc^{2} $ (in the laboratory frame) has been used, and $\gamma=\left(1-\frac{v^{2}}{c^{2}}\right)^{{\raise0.5ex\hbox{$\scriptstyle 1 $}\kern-0.1em/\kern-0.15em\lower0.25ex\hbox{$\scriptstyle 2 $}}}$ is the Lorentz transformation factor between the laboratory frame and the electron rest frame.

Consider now the state 

\begin{equation} \label{Eqn32} 
{\left| \sigma  \right\rangle} \equiv \frac{1}{2} \left({\left| E,\kappa ,m_{j}^{\sigma } ,+1 \right\rangle} +\sigma {\left| E,\kappa ,m_{j}^{\sigma } ,-1 \right\rangle} \right) 
\end{equation} 

\noindent where $m_{j}^{\sigma } \equiv \left(m_{\ell } +\frac{1}{2} \sigma \right)$.  By \eqref{Eqn28} and \eqref{Eqn29} we can write this as 

\begin{widetext}
\begin{equation} \label{Eqn33} 
{\left| \sigma  \right\rangle} =\left(\begin{array}{c} {\delta _{\sigma ,+} Z_{m_{\ell } } \left(\kappa \rho \right)e^{im_{\ell } \phi } } \\ {\delta _{\sigma ,-} Z_{m_{\ell } } \left(\kappa \rho \right)e^{im_{\ell } \phi } } \\ {\frac{\hbar c\left(\beta \delta _{\sigma ,+} Z_{m_{\ell } } \left(\kappa \rho \right)e^{im_{\ell } \phi } -i\kappa \delta _{\sigma ,-} Z_{m_{\ell } -1} \left(\kappa \rho \right)e^{i\left(m_{\ell } -1\right)\phi } \right)}{mc^{2} +E+eV_{0} } } \\ {\frac{\hbar c\left(i\kappa \delta _{\sigma ,+} Z_{m_{\ell } +1} \left(\kappa \rho \right)e^{i\left(m_{\ell } +1\right)\phi } -\beta \delta _{\sigma ,-} Z_{m_{\ell } } \left(\kappa \rho \right)e^{im_{\ell } \phi } \right)}{mc^{2} +E+eV_{0} } } \end{array}\right)e^{i\left(\beta z-\frac{E}{\hbar } t\right)}  
\end{equation} 
\end{widetext}

\noindent where $\delta _{\sigma ,\pm } $ are Kronecker delta functions and $\sigma =\pm 1$ as before.  We denote the state in \eqref{Eqn33} by ${\left| \sigma  \right\rangle} $ because it becomes an eigenstate of $\hat{S}_{z} =\frac{\hbar }{2} \Sigma$ with eigenvalue $\frac{\hbar }{2} \sigma $ in the paraxial regime where $\kappa <<\beta $.  Note also that in the same limit ${\left| \sigma  \right\rangle} $ is also an eigenstate of $\hat{L}_{z} =-i\hbar \frac{\partial }{\partial \phi } $ with eigenvalue $m_{\ell } $.

It will simplify the analysis considerably to boost to a frame in which the terms involving $\beta $ in both of the lower components of \eqref{Eqn33} become vanishingly small relative to those terms involving $\kappa $.  For an electron wave traveling with a sufficiently non-relativistic group velocity, such a frame will always exist provided that ${\tfrac{1}{2}} mv_{T}^{2} <<eV_{0} <<\left|\frac{\kappa }{\beta } \right|mc^{2} $, where the lower bound ensures the existence of bound states, while the upper bound constrains the potential energy in order to avoid pair creation, which would invalidate the single particle Dirac theory.  We henceforth assume that the above inequality holds and carry out the boost, so that in the new (barred) frame \eqref{Eqn33} is approximated as

\begin{equation} \label{Eqn34} 
{\left| \overline{\sigma}  \right\rangle} =e^{-\sigma {\tfrac{\alpha _{z} }{2}} } \left(\begin{array}{c} {\delta _{\sigma ,+} Z_{m_{\ell } } \left(\kappa \rho \right)e^{im_{\ell } \phi } } \\ {\delta _{\sigma ,-} Z_{m_{\ell } } \left(\kappa \rho \right)e^{im_{\ell } \phi } } \\ {\frac{-i\hbar c\kappa \delta _{\sigma ,-} Z_{m_{\ell } -1} \left(\kappa \rho \right)e^{i\left(m_{\ell } -1\right)\phi } }{2mc^{2} +\gamma _{z} eV_{0} } } \\ {\frac{i\hbar c\kappa \delta _{\sigma ,+} Z_{m_{\ell } +1} \left(\kappa \rho \right)e^{i\left(m_{\ell } +1\right)\phi } }{2mc^{2} +\gamma _{z} eV_{0} } } \end{array}\right)e^{-i\frac{E}{\hbar } t}  
\end{equation} 

\noindent where $\gamma _{z} $ is the Lorentz transformation factor between the laboratory frame and the barred frame such that $\gamma _{z} \approx \gamma $ since $\kappa <<\beta $, and where $\bar{V}_{0} =\gamma _{z} eV_{0} \approx eV_{0} $ and $\bar{E}\approx mc^{2} $ have been used.  

We now impose boundary conditions upon ${\left| \overline{\sigma}  \right\rangle} $ by requiring the wavefunctions to be finite at the origin, zero at infinity, and continuous across the step-potential $V\left(\rho \right)=V_{0} -\Delta V\Theta \left(\rho -a\right)$, similarly to section \ref{sec:sos}.  Note however that in the present case we drop the requirement of the existence of a continuous derivative of the wavefunction at the boundary.  The reason for this stems from the difference in order between the Schr\"{o}dinger and Dirac equations--the second order Schr\"{o}dinger equation requires \textit{two} conditions at the boundary (both continuity and a continuous derivative) in order to determine the wavefunction, while the first order Dirac equation requires only one. Application of these conditions on the boosted wavefunction \eqref{Eqn34} results in the characteristic equation 

\begin{align} \label{Eqn35} 
\frac{\kappa }{2mc^{2} +\gamma _{z} eV_{0} } & \frac{J_{m_{\ell } +\sigma } \left(\kappa a\right)}{J_{m_{\ell } } \left(\kappa a\right)} \nonumber \\ & = \frac{\tilde{\kappa }}{2mc^{2} +\gamma _{z} e\left(V_{0} -\Delta V\right)} \frac{K_{m_{\ell } +\sigma } \left(\tilde{\kappa }a\right)}{K_{m_{\ell } } \left(\tilde{\kappa }a\right)}  
\end{align} 

\noindent which, since $\gamma _{z} eV_{0} <<mc^{2} $, is well approximated by

\begin{equation} \label{Eqn36} 
\kappa \frac{J_{m_{\ell } +\sigma } \left(\kappa a\right)}{J_{m_{\ell } } \left(\kappa a\right)} -\tilde{\kappa }\frac{K_{m_{\ell } +\sigma } \left(\tilde{\kappa }a\right)}{K_{m_{\ell } } \left(\tilde{\kappa }a\right)} =\gamma _{z} \frac{e\Delta V}{2mc^{2} } \kappa \frac{J_{m_{\ell } +\sigma } \left(\kappa a\right)}{J_{m_{\ell } } \left(\kappa a\right)}  
\end{equation} 

\noindent In Appendix A we show that equation \eqref{Eqn36} is equivalent to the following condition:

\begin{align} \label{Eqn37} 
\kappa \frac{J_{\left|m_{\ell } \right|+1} \left(\kappa a\right)}{J_{\left|m_{\ell } \right|} \left(\kappa a\right)} & -\tilde{\kappa }\frac{K_{\left|m_{\ell } \right|+1} \left(\tilde{\kappa }a\right)}{K_{\left|m_{\ell } \right|} \left(\tilde{\kappa }a\right)} \nonumber \\ & =\sigma \frac{m_{\ell } }{\left|m_{\ell } \right|} \left(\gamma _{z} \frac{e\Delta V}{2mc^{2} } \right)\kappa \frac{J_{\left|m_{\ell } \right|+\sigma \frac{m_{\ell } }{\left|m_{\ell } \right|} } \left(\kappa a\right)}{J_{\left|m_{\ell } \right|} \left(\kappa a\right)}  
\end{align} 

\noindent Again, as in \eqref{Eqn22}, we have arrived at an equation for two unknowns $\kappa$ and $\tilde{\kappa }$, which is solved together with \eqref{Eqn31}. The solution for $\kappa$ then yields $\beta$ via \eqref{Eqn30}. 

 From \eqref{Eqn37} we can clearly see that $\kappa $ (and therefore also $\beta $) depends upon the quantity $\sigma \frac{m_{\ell } }{\left|m_{\ell } \right|} $, which we assumed in section \ref{sec:rotation} in order to arrive at the spin-dependent rotation effect of equation \eqref{Eqn14}.  In particular, $\sigma \frac{m_{\ell } }{\left|m_{\ell } \right|} =+1$ in \eqref{Eqn37} corresponds to the case of parallel spin and orbital angular momenta (with $\kappa \to \kappa ^{+} $), while $\sigma \frac{m_{\ell } }{\left|m_{\ell } \right|} =-1$ corresponds to anti-parallel angular momenta (with $\kappa \to \kappa ^{-} $).  In Appendix B, we show that equation \eqref{Eqn37} gives a prediction for the spatial wavefunction rotation rate $\delta \beta $ that agrees very well with that of equation \eqref{Eqn26}.  Therefore, we conclude that the Hermitian perturbation

\begin{equation} \label{Eqn37b}
\frac{ie\hbar}{4m^{2}c^{2}} \hat{S}\cdot \left(\nabla \times E\right)+\frac{e}{2m^{2} c^{2}} \hat{S}\cdot \left(E\times \hat{p}\right)
\end{equation}

\noindent in \eqref{Eqn15} is indeed the correct choice of Hamiltonian for the cylindrical spin-orbit interaction. For comparison, plots of the rotation rate $\Delta \beta$ as a function of $R\equiv\frac{2\pi a}{\lambda }\sqrt{\frac{2e\Delta V}{mc^{2}}}$ as given by both the Dirac and Foldy-Wouthuysen approaches are shown in Fig. 5.  In both plots we keep $eV=0.02mc^{2}$ constant, so that an increase in $R$ corresponds to an increase in the ratio of the cylinder potential radius to the Compton wavelength $\frac{a}{\lambda}$. While it can be seen from the figure that the plots from both approaches agree very well with one another, there is a small relative discrepancy which increases as $R$ becomes small, as higher order relativistic effects come into play. Furthermore, the predicted decrease in $\Delta \beta$ with increasing $R$ is to be expected, as the transverse electron wavefunctions will tunnel into the step-potential with decreasing amplitude as $\frac{a}{\lambda}$ increases. The two predictions also approach each other asymptotically in this regime, as expected.

\begin{figure}
\begin{center}
\includegraphics[width=0.5\textwidth]{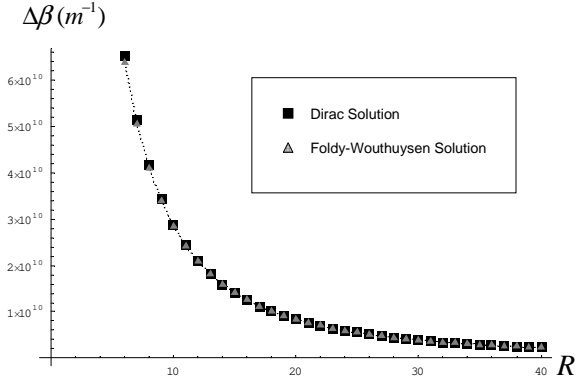}
\end{center}
\caption{\label{Fig5.eps} Plot of the Dirac and Foldy-Wouthuysen predictions for the rotation rate (propagation constant splitting) $\Delta \beta$ vs. $R\equiv\sqrt{2\left(\frac{2\pi a}{\lambda } \right)^{2} \left(\frac{e\Delta V}{mc^{2} } \right)}$, for $eV=0.02mc^{2}$.}
\end{figure}

 Having demonstrated the equivalence of the Dirac and Foldy-Wouthuysen approaches with regard to the cylindrical SOI phenomenon, our final aim is to derive the analogue of \eqref{Eqn14} in section \ref{sec:rotation}, showing the spatial rotation of the Dirac bispinors.  By an argument similar to that surrounding equation \eqref{Eqn14}, starting from the (non-boosted) equation \eqref{Eqn33} we find that in the paraxial regime $\kappa <<\beta $, for $\rho <a$, a \textit{parallel} bispinor (that is, $\sigma m_{\ell}=+1$) has the form

\begin{align} \label{Eqn38} 
{\left| \psi _{p}  \right\rangle} = & \left(\pm 1\right)^{\left|m_{\ell } \right|} \left(\begin{array}{c} {\delta _{\sigma ,+} } \\ {\delta _{\sigma ,-} } \\ {\frac{\hbar c\beta ^{+} }{mc^{2} +E+eV_{0} } \delta _{\sigma ,+} } \\ {\frac{-\hbar c\beta ^{+} }{mc^{2} +E+eV_{0} } \delta _{\sigma ,-} } \end{array}\right) \nonumber \\ & \times J_{\left|m_{\ell } \right|} \left(\kappa ^{+} \rho \right)e^{i\left(\sigma \left|m_{\ell } \right|\phi +\beta ^{+} z\right)} e^{-i\frac{E_{0} }{\hbar } t}  
\end{align} 

\noindent while an \textit{anti-parallel} bispinor ($\sigma m_{\ell}=-1$) is

\begin{align} \label{Eqn39} 
{\left| \psi _{ap}  \right\rangle} = & \left(\pm 1\right)^{\left|m_{\ell } \right|} \left(\begin{array}{c} {\delta _{\sigma ,+} } \\ {\delta _{\sigma ,-} } \\ {\frac{\hbar c\beta ^{-} }{mc^{2} +E+eV_{0} } \delta _{\sigma ,+} } \\ {\frac{-\hbar c\beta ^{-} }{mc^{2} +E+eV_{0} } \delta _{\sigma ,-} } \end{array}\right) \nonumber \\ & \times J_{\left|m_{\ell } \right|} \left(\kappa ^{-} \rho \right)e^{-i\left(\sigma \left|m_{\ell } \right|\phi -\beta ^{-} z\right)} e^{-i\frac{E_{0} }{\hbar } t}  
\end{align} 

The key point here is that we are working in the near-paraxial regime, where $\kappa ^{+} \approx \kappa ^{-} \approx \bar{\kappa }\equiv {\tfrac{1}{2}} \left(\kappa ^{+} +\kappa ^{-} \right)$ and $\beta ^{+} \approx \beta ^{-} =\bar{\beta }\equiv {\tfrac{1}{2}} \left(\beta ^{+} +\beta ^{-} \right)$.  This fact allows us to make the following approximation: we \textit{completely neglect }the small differences in transverse wavenumber $\kappa ^{\pm } $ and propagation constant $\beta ^{\pm } $ in the \textit{amplitudes} of the spinorial components of \eqref{Eqn38} and \eqref{Eqn39}, while \textit{retaining} the propagation constant differences in the phase factors $e^{i\beta ^{\pm } z} $.  This is a valid approximation, since a small varying phase difference between propagating superposition states can have a large qualitative effect on the evolution of the probability distribution, while small amplitude differences will have only a small effect on this evolution.  Under the aforementioned approximation, the approximate superposition ${\left| \psi _{\sigma }  \right\rangle} $ of \eqref{Eqn38} and \eqref{Eqn39} can be written as

\begin{align} \label{Eqn40} 
{\left| \psi _{\sigma }  \right\rangle} \approx & \left(\pm 1\right)^{\left|m_{\ell } \right|} \left(\begin{array}{c} {\delta _{\sigma ,+} } \\ {\delta _{\sigma ,-} } \\ {\frac{\hbar c\bar{\beta }}{mc^{2} +E+eV_{0} } \delta _{\sigma ,+} } \\ {\frac{-\hbar c\bar{\beta }}{mc^{2} +E+eV_{0} } \delta _{\sigma ,-} } \end{array}\right) J_{\left|m_{\ell } \right|} \left(\bar{\kappa }\rho \right) \nonumber \\ & \times \left[e^{i\left(\sigma \left|m_{\ell } \right|\phi +\beta ^{+} z\right)} +e^{-i\left(\sigma \left|m_{\ell } \right|\phi -\beta ^{-} z\right)} \right]e^{-i\frac{E_{0} }{\hbar } t}  
\end{align} 

\noindent Recalling from section \ref{sec:rotation} that $\left(e^{i\left(\sigma \left|m_{\ell } \right|\phi +\beta ^{+} z\right)} +e^{-i\left(\sigma \left|m_{\ell } \right|\phi -\beta ^{-} z\right)} \right)$ can be written as $\cos \left(\left|m_{\ell } \right|\phi +\sigma \Delta \beta z\right)e^{i\bar{\beta }z} $, we present the final form for the quasi-paraxial spin-dependent spatially rotating Dirac bispinors:

\begin{align} \label{Eqn41} 
{\left| \psi _{\sigma }  \right\rangle} = & \left(\pm 1\right)^{\left|m_{\ell } \right|} \left(\begin{array}{c} {\delta _{\sigma ,+} } \\ {\delta _{\sigma ,-} } \\ {\frac{\hbar c\bar{\beta }}{mc^{2} +E+eV_{0} } \delta _{\sigma ,+} } \\ {\frac{-\hbar c\bar{\beta }}{mc^{2} +E+eV_{0} } \delta _{\sigma ,-} } \end{array}\right)J_{\left|m_{\ell } \right|} \left(\bar{\kappa }\rho \right) \nonumber \\ & \times  \cos \left(\left|m_{\ell } \right|\phi +\sigma \Delta \beta z\right)e^{i\left(\bar{\beta }z-\frac{E}{\hbar } t\right)}  
\end{align} 

 If we neglect the two lower small components, we find that \eqref{Eqn41} does indeed approximately reduce to the two-component Schr\"{o}dinger spinor in \eqref{Eqn14}, and clearly shows the spatial rotation effect.

\section{\label{sec:conclusions}Conclusions\protect}

 We have shown via direct solution of the Dirac equation for a cylindrical step-potential that the SOI Hamiltonian derived heuristically as equation \eqref{Eqn5} and more rigorously as equation \eqref{Eqn20} correctly predicts a splitting of the dispersion curves of the electronic eigenstates according to the relative direction of their spin and orbital angular momenta. This splitting can cause a propagation constant (phase velocity) difference between parallel and anti-parallel states, which in turn gives rise to stable states that exhibit spin-controlled rotation of their spatial probability distributions. In particular, we found that for a given energy, a parallel electronic state has a slightly \textit{smaller} propagation constant than that of an anti-parallel state. Although we have treated only the simple case of a step-potential in detail, it is clear from \eqref{Eqn15b} that any inhomogeneous cylindrical potential that is translationally invariant in the z direction will give rise to a similar spin-orbit interaction.

 Another way of looking at the difference between parallel and anti-parallel states is the following: numerical solution of equation \eqref{Eqn37} implies that a parallel electronic state has a slightly \textit{larger} value for its transverse wavenumber $\kappa$ as compared to an anti-parallel state. It follows from this that the transverse radial wavefunction associated with a parallel state does \textit{not} penetrate as far into the step-potential as that associated with an anti-parallel state. 
 
 A similar SOI effect occurring for a photon propagating paraxially in a cylindrically symmetric step-index optical fiber can also be viewed in the above manner. For the photonic case, the step-index in the dielectric medium plays the role of the step-potential, and the photon helicity plays the role of the electron spin. Stable, spin-controlled, rotating photonic  superposition states with field distributions similar to those shown in Fig. 4 occur also for the photon case \cite{Liberman}, which arise from a similar splitting of the dispersion curves for parallel and anti-parallel photons. We note that the photonic spin-controlled rotational effect (called the optical Magnus effect) was predicted in \cite{Liberman} for a graded-index fiber with a parabolic profile and also for a step-index profile, however to our knowledge analytic results for the step-index case have not been presented in the context of the wave theory of SOI for a photon. 
 
 As is well known, the basis of the electronic SOI is the sum of two physical effects: the interaction of the electron's magnetic moment with the magnetic field resulting from the electron's motion through an inhomogeneous potential, and the Thomas precession resulting from the electron's curvilinear path of travel due to this potential. It is interesting, however, that for the analogous case of a photon propagating in an inhomogeneous medium, the SOI effect persists although the photon lacks a physical analogue to the electron's magnetic moment. The spin-orbit interaction of a particle with arbitrary spin has been discussed in \cite{Berard06}, in which the SOI is explained in terms of non-commutative space-time coordinates which arise from a non-Abelian Berry gauge connection.  In a future paper, we will give the details of the SOI calculation for photons in a step-index fiber, employing both the \textquotedblleft{perturbative}\textquotedblright $ $ and \textquotedblleft{exact}\textquotedblright $ $ approaches in parallel with this present work, in order to further elucidate the electron-photon SOI analogy.

\begin{acknowledgments}
We wish to acknowledge Steven van Enk for fruitful discussions. This work was supported by the National Science Foundation, through its Physics at the Information Frontier (PIF) program, grant PHY-0554842.
\end{acknowledgments}

\appendix
\section{Equivalent characteristic equation}

 It will be convenient in what follows to express the $K_{n} \left(x\right)$ functions (modified Bessel functions of the second kind with purely real arguments) of \eqref{Eqn36} in terms of $H_{n} \left(ix\right)$ (Hankel functions of the first kind with purely imaginary arguments). These functions are related by the identity \cite{TableofIntegrals2}

\begin{equation} \label{EqnA1} 
H_{n} \left(ix\right)=\left(-i\right)^{n+1} \frac{2}{\pi } K_{n} \left(x\right)
\end{equation}

\noindent The reason for this replacement is that the $J_{n} \left(x\right)$ and $H_{n} \left(ix\right)$ obey the same recursion relations \cite{TableofIntegrals3} while the $J_{n} \left(x\right)$ and $K_{n} \left(x\right)$ do not.  Using \eqref{EqnA1}, we find that equation \eqref{Eqn36} can be written as

\begin{equation} \label{EqnA2}
u\frac{J_{m_{\ell } +\sigma } \left(u\right)}{J_{m_{\ell } } \left(u\right)} -v\frac{H_{m_{\ell } +\sigma } \left(v\right)}{H_{m_{\ell } } \left(v\right)} =\varepsilon u\frac{J_{m_{\ell } +\sigma } \left(u\right)}{J_{m_{\ell } } \left(u\right)}
\end{equation}

\noindent where $u\equiv \kappa a$, $v\equiv i\tilde{\kappa }a$, $\varepsilon \equiv \gamma _{z} \frac{e\Delta V}{2mc^{2} } $, and where we have for later convenience multiplied both sides by $a$.  

 Our next task is to explicitly account for the absolute sign of $m_{\ell } $ in equation \eqref{EqnA2}.  We therefore replace $m_{\ell } \to \pm \left|m_{\ell } \right|$ in \eqref{EqnA2}, where the upper sign corresponds to case where $m_{\ell } >0$ (positive OAM) in the spinor in \eqref{Eqn33}, while the lower sign corresponds to $m_{\ell } <0$ (negative OAM).  Under this replacement, \eqref{EqnA2} becomes

\begin{equation} \label{EqnA3}
u\frac{J_{\pm \left|m_{\ell } \right|+\sigma } \left(u\right)}{J_{\pm \left|m_{\ell } \right|} \left(u\right)} -v\frac{H_{\pm \left|m_{\ell } \right|+\sigma } \left(v\right)}{H_{\pm \left|m_{\ell } \right|} \left(v\right)} =\varepsilon u\frac{J_{\pm \left|m_{\ell } \right|+\sigma } \left(u\right)}{J_{\pm \left|m_{\ell } \right|} \left(u\right)}
\end{equation}

\noindent We can re-express this, however, using the Bessel function relations \cite{TableofIntegrals4} $Z_{-n} \left(x\right)=\left(-1\right)^{n} Z_{n} \left(x\right)$ (the $Z$-functions stand for either the $J_{n} \left(x\right)$ or the $H_{n} \left(x\right)$ cylinder functions), so that \eqref{EqnA3} becomes

\begin{equation} \label{EqnA4}
\noindent u\frac{J_{\left|m_{\ell } \right|\pm \sigma } \left(u\right)}{J_{\left|m_{\ell } \right|} \left(u\right)} -v\frac{H_{\left|m_{\ell } \right|\pm \sigma } \left(v\right)}{H_{\left|m_{\ell } \right|} \left(v\right)} =\varepsilon u\frac{J_{\left|m_{\ell } \right|\pm \sigma } \left(u\right)}{J_{\left|m_{\ell } \right|} \left(u\right)}
\end{equation}

Now, for $\sigma =+1$ \eqref{EqnA4} becomes 

\begin{equation} \label{EqnA5}
u\frac{J_{\left|m_{\ell } \right|\pm 1} \left(u\right)}{J_{\left|m_{\ell } \right|} \left(u\right)} -v\frac{H_{\left|m_{\ell } \right|\pm 1} \left(v\right)}{H_{\left|m_{\ell } \right|} \left(v\right)} =\varepsilon u\frac{J_{\left|m_{\ell } \right|\pm 1} \left(u\right)}{J_{\left|m_{\ell } \right|} \left(u\right)}
\end{equation}

\noindent which consists of two distinct equations, corresponding to the choice of sign in the expression $\left|m_{\ell } \right|\pm 1$ appearing in the cylinder function arguments in the numerators.  Focusing now on the case involving $\left|m_{\ell } \right|-1$, where the left hand side of \eqref{EqnA5} is

\begin{equation} \label{EqnA6}
u\frac{J_{\left|m_{\ell } \right|-1} \left(u\right)}{J_{\left|m_{\ell } \right|} \left(u\right)} -v\frac{H_{\left|m_{\ell } \right|-1} \left(v\right)}{H_{\left|m_{\ell } \right|} \left(v\right)}
\end{equation}

\noindent we employ the following Bessel function identity 

\begin{align} \label{EqnA7} 
u\frac{J_{\left|m_{\ell } \right|-1} \left(u\right)}{J_{\left|m_{\ell } \right|} \left(u\right)} & -v\frac{H_{\left|m_{\ell } \right|-1} \left(v\right)}{H_{\left|m_{\ell } \right|} \left(v\right)} \nonumber \\ & =-u\frac{J_{\left|m_{\ell } \right|+1} \left(u\right)}{J_{\left|m_{\ell } \right|} \left(u\right)} -v\frac{H_{\left|m_{\ell } \right|+1} \left(v\right)}{H_{\left|m_{\ell } \right|} \left(v\right)}
\end{align}

\noindent which can be proved by substituting the fundamental identities \cite{TableofIntegrals5} $Z_{n-1} \left(x\right)=\frac{2n}{x} Z_{n} \left(x\right)-Z_{n+1} \left(x\right)$ (again, $Z$ stands for either $J_{n} \left(x\right)$ or $H_{n} \left(x\right)$) into the left hand side of \eqref{EqnA7}.  Using \eqref{EqnA7} in \eqref{EqnA6} allows us to write the two equations in \eqref{EqnA5} as

\begin{equation} \label{EqnA8}
u\frac{J_{\left|m_{\ell } \right|+1} \left(u\right)}{J_{\left|m_{\ell } \right|} \left(u\right)} -v\frac{H_{\left|m_{\ell } \right|+1} \left(v\right)}{H_{\left|m_{\ell } \right|} \left(v\right)} =\pm \varepsilon u\frac{J_{\left|m_{\ell } \right|\pm 1} \left(u\right)}{J_{\left|m_{\ell } \right|} \left(u\right)}
\end{equation}
\noindent 

\noindent We note here that \eqref{EqnA8} is an expression of \eqref{EqnA4} in the case where $\sigma =+1$ (note also the new factor of $\pm 1$ on the right hand side of \eqref{EqnA8}).

 We now insert $\sigma =-1$ into \eqref{EqnA4}, and carry out a simplification analogous to \eqref{EqnA5}-\eqref{EqnA8}, concluding that for $\sigma =-1$, \eqref{EqnA4} is equivalent to 

\begin{equation} \label{EqnA9}
u\frac{J_{\left|m_{\ell } \right|+1} \left(u\right)}{J_{\left|m_{\ell } \right|} \left(u\right)} -v\frac{H_{\left|m_{\ell } \right|+1} \left(v\right)}{H_{\left|m_{\ell } \right|} \left(v\right)} =\mp \varepsilon u\frac{J_{\left|m_{\ell } \right|\mp 1} \left(u\right)}{J_{\left|m_{\ell } \right|} \left(u\right)}
\end{equation}

\noindent Comparing \eqref{EqnA8} and \eqref{EqnA9}, it is apparent that both equations can be expressed simultaneously via

\begin{equation} \label{EqnA10}
 u\frac{J_{\left|m_{\ell } \right|+1} \left(u\right)}{J_{\left|m_{\ell } \right|} \left(u\right)} -v\frac{H_{\left|m_{\ell } \right|+1} \left(v\right)}{H_{\left|m_{\ell } \right|} \left(v\right)} =\pm \sigma \varepsilon u\frac{J_{\left|m_{\ell } \right|\pm \sigma } \left(u\right)}{J_{\left|m_{\ell } \right|} \left(u\right)}
\end{equation}

\noindent Recalling that the upper and lower signs in \eqref{EqnA10} denote the cases of positive and negative $m_{\ell } $ respectively, we replace the ``$\pm $'' notation on the right hand side with the equivalent expression $\frac{m_{\ell } }{\left|m_{\ell } \right|} $.  In this way, we arrive at our final form for the characteristic equation,

\begin{equation} \label{EqnA11}
u\frac{J_{\left|m_{\ell } \right|+1} \left(u\right)}{J_{\left|m_{\ell } \right|} \left(u\right)} -v\frac{H_{\left|m_{\ell } \right|+1} \left(v\right)}{H_{\left|m_{\ell } \right|} \left(v\right)} =\sigma \frac{m_{\ell } }{\left|m_{\ell } \right|} \varepsilon u\frac{J_{\left|m_{\ell } \right|+\sigma \frac{m_{\ell } }{\left|m_{\ell } \right|} } \left(u\right)}{J_{\left|m_{\ell } \right|} \left(u\right)}
\end{equation}

\noindent which, in light of (1A), is just equation \eqref{Eqn37} of section \ref{sec:sos}.

\section{Wavefunction rotation rate}

Our aim here is to expand equation \eqref{Eqn37} in such a way that it gives rise to an equation of the form \eqref{Eqn26} for the wavefunction rotation rate $\Delta \beta $, thus showing that the exact solution presented in section \ref{sec:sos} is well approximated by the perturbative approach of section \ref{sec:rotation}.  We start therefore with equation \eqref{Eqn37}, written in the form given in appendix A as equation \eqref{EqnA11},

\begin{equation} \label{EqnB1}
u\frac{J_{\left|m_{\ell } \right|+1} \left(u\right)}{J_{\left|m_{\ell } \right|} \left(u\right)} -v\frac{H_{\left|m_{\ell } \right|+1} \left(v\right)}{H_{\left|m_{\ell } \right|} \left(v\right)} =\sigma \frac{m_{\ell } }{\left|m_{\ell } \right|} \varepsilon u\frac{J_{\left|m_{\ell } \right|+\sigma \frac{m_{\ell } }{\left|m_{\ell } \right|} } \left(u\right)}{J_{\left|m_{\ell } \right|} \left(u\right)} 
\end{equation}

\noindent recalling that $u\to u^{+} \equiv \kappa ^{+} a$ and $v\to v^{+} \equiv i\tilde{\kappa }^{+} a$ or $u\to u^{-} \equiv \kappa ^{-} a$ and $v\to v^{-} \equiv i\tilde{\kappa }^{-} a$

\noindent depending on whether $\sigma \frac{m_{\ell } }{\left|m_{\ell } \right|} =+1$ or $\sigma \frac{m_{\ell } }{\left|m_{\ell } \right|} =-1$, respectively.  Since $\varepsilon \equiv \gamma _{z} \frac{e\Delta V}{2mc^{2} } $ is small, however, we have from \eqref{EqnB1} that $u^{+} \approx u^{-} $.  We can exploit this by adding/subtracting equation \eqref{EqnB1} with $\sigma \frac{m_{\ell } }{\left|m_{\ell } \right|} =+1$ to/from equation \eqref{EqnB1} with $\sigma \frac{m_{\ell } }{\left|m_{\ell } \right|} =-1$.

Adding the two cases of this equation gives

\begin{align} \label{EqnB2}
u^{+} \Im \left(u^{+} \right) &+ u^{-} \Im \left(u^{-} \right)-v^{+} \aleph \left(v^{+} \right)-v^{-} \aleph \left(v^{-} \right) \nonumber \\
& \approx \varepsilon \left[u^{+} \frac{J_{\left|m_{\ell } \right|+1} \left(u^{+} \right)}{J_{\left|m_{\ell } \right|} \left(u^{+} \right)} -u^{-} \frac{J_{\left|m_{\ell } \right|-1} \left(u^{-} \right)}{J_{\left|m_{\ell } \right|} \left(u^{-} \right)} \right]
\end{align}

\noindent where $\Im \left(x\right)\equiv \frac{J_{\left|m_{\ell } \right|+1} \left(x\right)}{J_{\left|m_{\ell } \right|} \left(x\right)}$, $\aleph \left(x\right)\equiv \frac{H_{\left|m_{\ell } \right|+1} \left(x\right)}{H_{\left|m_{\ell } \right|} \left(x\right)} $.  We now multiply both sides of \eqref{EqnB2} by ${\tfrac{a}{2}} $ and Taylor expand $\Im \left(u^{\pm } \right)$ and $\frac{J_{\left|m_{\ell } \right|\pm 1} \left(u^{\pm } \right)}{J_{\left|m_{\ell } \right|} \left(u^{\pm } \right)} $ about the point $\bar{u}\equiv {\tfrac{1}{2}} \left(u^{+} +u^{-} \right)$, to first order in $\delta u$, where $\delta u\equiv {\tfrac{1}{2}} \left(u^{+} -u^{-} \right)$ so that $u^{\pm } =\bar{u}\pm \delta u$; we also Taylor expand $\aleph \left(v\right)$ about the point $\bar{v}\equiv {\tfrac{1}{2}} \left(v^{+} +v^{-} \right)$, to first order in $\delta v$, where $\delta v\equiv {\tfrac{1}{2}} \left(v^{+} -v^{-} \right)$ so that $v^{\pm } =\bar{v}\pm \delta v$.  These substitutions result in the following equation:

\begin{widetext}
\begin{align} \label{EqnB3}
\bar{u}\Im \left(\bar{u}\right)-\bar{v}\aleph \left(\bar{v}\right) + \left\{\left(\delta u\right)^{2} \Im'\left(\bar{u}\right) -\left(\delta v\right)^{2} \aleph '\left(\bar{v}\right)\right\} & \approx \varepsilon \bar{u}\frac{J_{\left|m_{\ell } \right|+1} \left(\bar{u}\right)-J_{\left|m_{\ell } \right|-1} \left(\bar{u}\right)}{J_{\left|m_{\ell } \right|} \left(\bar{u}\right)} +\varepsilon \delta u \Bigg\{ \frac{J_{\left|m_{\ell } \right|+1} \left(\bar{u}\right)+J_{\left|m_{\ell } \right|-1} \left(\bar{u}\right)}{J_{\left|m_{\ell } \right|} \left(\bar{u}\right)} \nonumber \\ 
& +\bar{u}\frac{\left(J'_{\left|m_{\ell } \right|+1} \left(\bar{u}\right)+J'_{\left|m_{\ell } \right|-1} \left(\bar{u}\right)\right)J_{\left|m_{\ell } \right|} \left(\bar{u}\right)-\left(J_{\left|m_{\ell } \right|+1} \left(\bar{u}\right)+J_{\left|m_{\ell } \right|-1} \left(\bar{u}\right)\right)}{2J_{\left|m_{\ell } \right|}^{2} \left(\bar{u}\right)} \Bigg\}
\end{align} 
\end{widetext}

\noindent where the primes denote derivatives with respect to functional arguments.  In (B3), the term in curly brackets on the left hand side is negligible because it is second order in $\left(\delta u\right)^{2} $ and $\left(\delta v\right)^{2} $, while the term in curly brackets on the right hand side is negligible because both $\varepsilon $ and $\delta u$ are small quantities.  Thus, to first order we have

\begin{equation} \label{EqnB4}
\bar{u}\Im \left(\bar{u}\right)-\bar{v}\aleph \left(\bar{v}\right)=\varepsilon \bar{u}\frac{J_{\left|m_{\ell } \right|+1} \left(\bar{u}\right)-J_{\left|m_{\ell } \right|-1} \left(\bar{u}\right)}{J_{\left|m_{\ell } \right|} \left(\bar{u}\right)} \end{equation}

\noindent and since \eqref{Eqn31} implies that 

\begin{equation} \label{EqnB5}
\bar{v}=i\sqrt{R_{\gamma }^{2} -\bar{u}^{2} }
\end{equation}

\noindent we have in \eqref{EqnB4} and \eqref{EqnB5} two equations in the two unknown variables $\bar{u}$ and $\bar{v}$.  
Therefore, upon substituting for $\bar{v}$ via \eqref{EqnB5}, \eqref{EqnB4} can be solved for $\bar{u}$ numerically.

 We now \textit{subtract} the equations \eqref{EqnB1}, and find 

\begin{align} \label{EqnB8}
u^{+} \Im \left(u^{+} \right) & -  u^{-} \Im \left(u^{-} \right)- \left(v^{+} \aleph \left(v^{+} \right)-v^{-} \aleph \left(v^{-} \right)\right)  \nonumber \\ & =\varepsilon \left[u^{+} \frac{J_{\left|m_{\ell } \right|+1} \left(u^{+} \right)}{J_{\left|m_{\ell } \right|} \left(u^{+} \right)} +u^{-} \frac{J_{\left|m_{\ell } \right|-1} \left(u^{-} \right)}{J_{\left|m_{\ell } \right|} \left(u^{-} \right)} \right]
\end{align}

\noindent Multiplying \eqref{EqnB8} by ${\tfrac{a}{2}} $ as before, again Taylor expanding to first order in $\delta u$ and $\delta v$ about $\bar{u}$ and $\bar{v}$, and neglecting quantities of order $\left(\delta u\right)^{2} $ and $\varepsilon \delta u$, we arrive at

\begin{align} \label{EqnB9}
\delta u\Im \left(\bar{u}\right)+\bar{u}\Im '\left(\bar{u}\right)\delta u & -  \left(\delta v\aleph \left(\bar{v}\right)+\bar{v}\aleph '\left(\bar{v}\right)\delta v\right) \nonumber \\ & \approx \frac{\varepsilon }{2} \left(\bar{u}\frac{J_{\left|m_{\ell } \right|+1} \left(\bar{u}\right)+J_{\left|m_{\ell } \right|-1} \left(\bar{u}\right)}{J_{\left|m_{\ell } \right|} \left(\bar{u}\right)} \right)
\end{align}

\noindent Now, using \eqref{Eqn31} to expand $v$ to first order, we find that 

\begin{align} \label{EqnB10}
v\approx \sqrt{\bar{u}^{2} -R_{\gamma }^{2} } \pm \frac{\bar{u}\delta u}{\sqrt{\bar{u}^{2} -R_{\gamma }^{2} } } =\bar{v}\pm \delta v
\end{align}

\noindent so that 

\begin{equation} \label{EqnB11}
\delta v=\frac{\bar{u}\delta u}{\sqrt{\bar{u}^{2} -R_{\gamma }^{2} } } =\frac{\bar{u}}{\bar{v}} \delta u
\end{equation}

\noindent which can be used in \eqref{EqnB9} to yield

\begin{align} \label{EqnB12}
\delta u\frac{1}{\bar{u}} \Im \left(\bar{u}\right)+\Im '\left(\bar{u}\right) & -  \frac{1}{\bar{v}} \aleph \left(\bar{v}\right)+\aleph '\left(\bar{v}\right) \nonumber \\ & \approx \frac{\varepsilon }{2} \left(\frac{J_{\left|m_{\ell } \right|+1} \left(\bar{u}\right)+J_{\left|m_{\ell } \right|-1} \left(\bar{u}\right)}{J_{\left|m_{\ell } \right|} \left(\bar{u}\right)} \right)
\end{align}

 To simplify the term on the left hand side, we substitute $\Im \left(x\right)\equiv \frac{J_{\left|m_{\ell } \right|+1} \left(x\right)}{J_{\left|m_{\ell } \right|} \left(x\right)} $ and  $\aleph \left(x\right)\equiv \frac{H_{\left|m_{\ell } \right|+1} \left(x\right)}{H_{\left|m_{\ell } \right|} \left(x\right)} $, while again using (1A) and (5B) along with the cylinder function relations \cite{TableofIntegrals6} $Z'_{n} \left(x\right)=Z_{n-1} \left(x\right)-\frac{n}{x} Z_{n+1} \left(x\right)$ and $Z'_{n+1} \left(x\right)=Z_{n} \left(x\right)-\frac{n+1}{x} Z_{n+1} \left(x\right)$ ($Z$ stands for either $J_{n} \left(x\right)$ or $H_{n} \left(x\right)$), in order to obtain

\begin{align} \label{EqnB13} 
 \frac{1}{\bar{u}} & \Im \left(\bar{u}\right)+\Im '\left(\bar{u}\right)-\frac{1}{\bar{v}} \aleph \left(\bar{v}\right)+\aleph '\left(\bar{v}\right)= \nonumber \\ & \frac{K_{\left|m_{\ell } \right|-1} \left(-i\bar{v}\right)K_{\left|m_{\ell } \right|+1} \left(-i\bar{v} \right)}{K_{\left|m_{\ell } \right|}^{2} \left(-i\bar{v}\right)} -\frac{J_{\left|m_{\ell } \right|-1} \left(\bar{u}\right)J_{\left|m_{\ell } \right|+1} \left(\bar{u}\right)}{J_{\left|m_{\ell } \right|}^{2} \left(\bar{u}\right)}
\end{align}

\noindent For the term on the right hand side of \eqref{EqnB12}, we use \cite{TableofIntegrals5} $\frac{2n}{x} J_{n} \left(x\right)=J_{n+1} \left(x\right)+J_{n-1} \left(x\right)$ so that

\begin{align} \label{EqnB14}
\frac{\varepsilon }{2} \left(\frac{J_{\left|m_{\ell } \right|+1} \left(\bar{u}\right)+J_{\left|m_{\ell } \right|-1} \left(\bar{u}\right)}{J_{\left|m_{\ell } \right|} \left(\bar{u}\right)} \right) \approx \varepsilon \frac{\left|m_{\ell } \right|}{\bar{u}}
\end{align}

Substituting the results \eqref{EqnB13} and \eqref{EqnB14} in \eqref{EqnB12} and solving for $\bar{u}\delta u$ then gives

\begin{align} \label{EqnB15}
\bar{u}\delta u \approx \varepsilon \left|m_{\ell } \right| & \Bigg[\frac{K_{\left|m_{\ell } \right|-1} \left(-i\bar{v}\right)K_{\left|m_{\ell } \right|+1} \left(-i\bar{v}\right)}{K_{\left|m_{\ell } \right|}^{2} \left(-i\bar{v}\right)} \nonumber \\ 
& -\frac{J_{\left|m_{\ell } \right|-1} \left(\bar{u}\right)J_{\left|m_{\ell } \right|+1} \left(\bar{u}\right)}{J_{\left|m_{\ell } \right|}^{2} \left(\bar{u}\right)} \Bigg]^{-1} 
\end{align}

\noindent $\delta u$ can thereby be found by substitution of $\bar{u}$ and $\bar{v}$ as given numerically by \eqref{EqnB4} and \eqref{EqnB5}.

 Having calculated the difference between the transverse wavenumbers $\delta u\equiv {\tfrac{1}{2}} a\left(\kappa ^{+} -\kappa ^{-} \right)$, we can now find the difference between the associated propagation constants, $\Delta \beta \equiv {\tfrac{1}{2}} \left(\beta ^{+} -\beta ^{-} \right)$. To do this we start with equation \eqref{Eqn30}, which upon substitution of $E\approx \gamma _{z} mc^{2} $  is equivalent to

\begin{align} \label{EqnB17}
\left(\beta ^{\pm } \right)^{2} \approx \frac{1}{c^{2}\hbar^{2} } \Big[ & \left(\gamma _{z}^{2} - 1\right)m^{2} c^{4} + 2\gamma _{z} mc^{2} eV \nonumber \\ 
& +\left(eV\left(\rho \right)\right)^{2}-c^{2} \hbar ^{2} \left(\kappa ^{\pm } \right)^{2} \Big]
\end{align}

\noindent in the unbarred (laboratory) frame.  Taking the square root of both sides and using $\left(\gamma _{z}^{2} -1\right)=\left(\gamma _{z} \frac{v_{z} }{c} \right)^{2} $ and $\gamma _{z} eV_{0} <<mc^{2} $ in order to Taylor expand the radical then gives

\begin{align} \label{EqnB19}
\beta ^{\pm } \approx \gamma _{z} \frac{mv_{z} }{\hbar } \Bigg[1-\frac{eV}{\gamma _{z} mv_{z}^{2} } + \frac{1}{2} & \left(\frac{c}{v_{z} } \frac{eV}{\gamma _{z} mc^{2} } \right)^{2} \nonumber \\
& -\frac{1}{2} \left(\frac{\hbar \kappa ^{\pm } }{\gamma _{z} mv_{z} } \right)^{2} \Bigg]
\end{align}

\noindent Therefore, we find that

\begin{equation} \label{EqnB20}
\Delta \beta =-\frac{1}{4} \frac{\hbar }{\gamma _{z} mv_{z} } \left[\left(\kappa ^{+} \right)^{2} -\left(\kappa ^{-} \right)^{2} \right]
\end{equation}

\noindent However, since $\kappa ^{\pm } a=\bar{u}\pm \delta u$, to first order in $\delta u$ \eqref{EqnB20} is equivalent to 

\begin{equation} \label{EqnB21}
\Delta \beta =-\frac{\hbar }{\gamma _{z} mv_{z} } \frac{1}{a^{2} } \bar{u}\delta u
\end{equation}

\noindent Substituting \eqref{EqnB15} into \eqref{EqnB21} then yields the desired expression for $\Delta \beta $,

\begin{align} \label{EqnB22}
\Delta \beta  = -\left|m_{\ell } \right| \frac{1}{v_{z} } \frac{\hbar e\Delta V}{2m^{2} c^{2} a^{2} } & \Bigg\{  \frac{K_{\left|m_{\ell } \right|-1} \left(-i\bar{v}\right)K_{\left|m_{\ell } \right|+1} \left(-i\bar{v}\right)}{K_{\left|m_{\ell } \right|}^{2} \left(-i\bar{v}\right)} \nonumber \\
& -\frac{J_{\left|m_{\ell } \right|-1} \left(\bar{u}\right)J_{\left|m_{\ell } \right|+1} \left(\bar{u}\right)}{J_{\left|m_{\ell } \right|}^{2} \left(\bar{u}\right)} \Bigg\}^{-1}  
\end{align}

\noindent where $\varepsilon \equiv \gamma _{z} \frac{e\Delta V}{2mc^{2} } $ has been used.  For clarity, we remind the reader that $-i\bar{v}=\sqrt{R_{\gamma }^{2} -\bar{u}^{2} }$ via \eqref{EqnB5}, where $R_{\gamma }^{2} $ is defined in equation \eqref{Eqn31}, and that $\bar{u}$ is found by solving equation \eqref{EqnB4}.

 Comparing this first-order result of \eqref{EqnB22} to equation \eqref{Eqn26}, we see that the two equations are of the same form.  Furthermore, since $\varepsilon <<1$ and $\gamma _{z} \approx 1$, equation \eqref{EqnB4} is nearly equivalent to equation \eqref{Eqn22}, so that $\bar{u}\approx \kappa _{0}a$ and $\sqrt{R_{\gamma }^{2} -\bar{u}^{2} } \approx \tilde{\kappa }_{0} a$.  We have therefore demonstrated that the perturbative approach of section \ref{sec:sos} is approximately equivalent to the direct approach of section \ref{sec:dirac}.

\end{document}